\documentclass[preprint]{revtex4}
\usepackage{latexsym}
\usepackage{amsfonts}
\usepackage{graphicx}
\usepackage{epsfig}
\usepackage{color}
\usepackage{comment}
\usepackage{subfigure}

\usepackage{setspace} 

\usepackage{amsmath}

 2

\begin{document}

\bigskip
\bigskip
\centerline{\bf Role of electrostatic interactions in amyloid $\beta$-protein (A$\beta$) oligomer formation:}
\centerline{\bf A discrete molecular dynamics study}

\bigskip

\noindent{\scriptsize
S.~Yun$^{1 \ast}$, B.~Urbanc$^{1 \ast}$, L.~Cruz$^1$,
G.~Bitan$^2$, D.~B.~Teplow$^2$, and
H.~E.~Stanley$^1$ } \medskip

\noindent
{\scriptsize
$^1$Center for Polymer Studies, Department of Physics, Boston
University, Boston, MA 02215 \\
$^2$Department of Neurology, David Geffen School of Medicine,
Brain Research Institute and Molecular Biology Institute, University of California, Los Angeles, CA 90095}

\bigskip
\bigskip
\noindent
$^\ast$Corresponding authors \\

\bigskip
\noindent
Corresponding authors.  E-mail: sijung@physics.bu.edu, brigita@bu.edu \\


\bigskip 
\noindent
{\bf Abbreviations:} AD, Alzheimer's disease; A$\beta$,
amyloid $\beta$-protein; DMD, discrete molecular dynamics; EIs, electrostatic interactions.

\newpage
\begin{center}
{\bf ABSTRACT}
\end{center}
\noindent 
Pathological folding and oligomer formation of the amyloid
$\beta$-protein (A$\beta$) are widely perceived as central to
Alzheimer's disease (AD).  Experimental approaches to study
A$\beta$ self-assembly provide limited information because
most relevant aggregates are quasi-stable and inhomogeneous. 
We apply a discrete molecular dynamics (DMD) approach
combined with a four-bead protein model to study oligomer
formation of A$\beta$.  We address the differences between
the two most common A$\beta$ alloforms, A$\beta$40 and
A$\beta$42, which oligomerize differently {\it in vitro}. 
Our previous study showed that, despite simplifications, our
DMD approach accounts for the experimentally observed
differences between A$\beta$40 and A$\beta$42 and yields
structural predictions amenable to {\it in vitro} testing. 
Here we study how the presence of electrostatic interactions
(EIs) between pairs of charged amino acids affects A$\beta$40
and A$\beta$42 oligomer formation.  Our results indicate that
EIs promote formation of larger oligomers in both A$\beta$40
and A$\beta$42.  Both A$\beta$40 and A$\beta$42 display a
peak at trimers/tetramers, but A$\beta$42 displays additional
peaks at nonamers and tetradecamers.  EIs thus shift the
oligomer size distributions to larger oligomers. 
Nonetheless, the A$\beta$40 size distribution remains
unimodal, whereas the A$\beta$42 distribution is trimodal, as
observed experimentally.  We show that structural differences
between A$\beta$40 and A$\beta$42 that already appear in the
monomer folding, are not affected by EIs.  A$\beta$42 folded
structure is characterized by a turn in the C-terminus that
is not present in A$\beta$40.  We show that the same
C-terminal region is also responsible for the strongest
intermolecular contacts in A$\beta$42 pentamers and larger
oligomers.  Our results suggest that this C-terminal region
plays a key role in the formation of A$\beta$42 oligomers and
the relative importance of this region increases in the
presence of EIs.  These results suggest that inhibitors
targeting the C-terminal region of A$\beta$42 oligomers may
be able to prevent oligomer formation or structurally modify
the assemblies to reduce their toxicity.


\newpage
\section*{Introduction}

Alzheimer's disease (AD) is a progressive brain disorder,
clinically characterized by the accumulation of extracellular
amyloid deposits composed of amyloid $\beta$-protein
(A$\beta$), intracellular neurofibrillary tangles, and
neuronal loss.  Recent research supports the hypothesis that
cerebral A$\beta$ accumulation is the primary cause of
neurotoxicity in AD~\citep{Hardy_02_Science}.  Accumulating
evidence suggests that A$\beta$ oligomers and prefibrillar
aggregates are the proximal effectors of neurotoxicity in the
early stages of AD~\citep{Kirkitadze_02_JNR,Klein_04_NA}. 
The predominant forms of A$\beta$ found in brains of AD
patients are 40 amino acids long (A$\beta$40) and 42 amino
acids long (A$\beta$42).  A$\beta$42 is linked particularly
strongly with AD.  Genetic studies have shown that autosomal
dominant forms of AD invariably involve increased production
of A$\beta$ or an increased A$\beta$42/A$\beta$40
concentration ratio~\citep{Sawamura_00_JBC}.  A$\beta$42
forms fibrils at significantly higher rates than does
A$\beta$40~\citep{Jarrett_93_Biochemistry,Jarrett_93_AnnNYAcadSci}
and A$\beta$42 self-association produces structures that are
more neurotoxic than those formed by
A$\beta$40~\citep{Klein_04_NA}.  Experimentally, there is a
distinct difference in oligomerization pathways of A$\beta$40
and A$\beta$42~\citep{Bitan_03_PNAS}.  \textit{In vitro}
experiment using the techniques, photo-induced cross-linking
of unmodified proteins (PICUP), size-exclusion
chromatography, dynamic light scattering, circular dichroism
spectroscopy, and electron microscopy showed that A$\beta$
exists as monomers, dimers, trimers, tetramers, and larger
oligomers in rapid equilibrium.  The A$\beta$40 oligomer size
distribution comprises monomer, dimer, trimer, and tetramer,
in similar amounts, and few higher-order oligomers.  The
A$\beta$42 distribution is multimodal, displaying a prominent
peak of pentamers/hexamers and smaller peaks of dodecamers
and octadecamers~\citep{Bitan_03_PNAS}.

Detailed, quantitative analysis of the three-dimensional
structures, energetics, and dynamics of oligomer formation
are necessary steps toward a molecular understanding of
A$\beta$ assembly and neurotoxicity.  During the formation of
fibrils, oligomers of different sizes co-exist with monomers
and larger aggregates such as
protofibrils~\citep{Walsh_97_JBC} and fibrils.  The relative
amounts of each oligomer type are small, which makes
determination of the structural properties of the oligomers
difficult.  Computer simulations, in contrast, are not
subject to the same kinds of problems, allowing small
oligomers to be studied at full atomic resolution (for recent
reviews, see \citep{Teplow_06_ACR}, \citep{Urbanc_06_CAR},
and \citep{Urbanc_06_ME}).

Conventional ``all-atom'' molecular dynamics (all-atom MD)
simulations with explicit solvent which take account of all
the protein and solvent atoms give the most detailed
information.  However, aggregation studies using all-atom MD
with explicit solvent are currently limited to either
aggregation of small number of A$\beta$ fragments such as
three A$\beta$(16-22) peptides~\citep{Klimov_03_Structure} or
stability studies of various A$\beta$ dimers with
predetermined structures~\citep{Tarus_05_JMB,Huet_06_BJ}. 
Tarus {\it et al.} used a protocol based on shape
complementarity to determine the initial A$\beta_{10-35}$
dimer structure and showed that the peptide dimers are
stabilized primarily through hydrophobic
interactions~\citep{Tarus_05_JMB}.  Huet {\it et al.} studied
A$\beta$40 and A$\beta$42 dimers and their A21G conformers,
starting from their fibrillar conformations and found various
possible topologies of dimers in
equilibrium~\citep{Huet_06_BJ}.  Keeping track of positions
and velocities of all the atoms at every time step is
computationally expensive.  Consequently, the times simulated
by the all-atom MD simulations are limited to few
microseconds~\citep{Urbanc_06_CAR}.  However, protein folding
and aggregation usually occur on time scales larger than
milliseconds.  To overcome this limitation, we use fast and
efficient discrete molecular dynamics (DMD)
simulations~\citep{Rapaport_97_Book} with a simplified
four-bead protein model and implicit solvent.  DMD is a
simplified version of MD using combination of square-well
potentials.  The DMD approach with a simplified protein model
and implicit solvent increases the efficiency of protein
folding and aggregation studies by a factor of $\sim$10$^7$
compared to the all-atom MD~\citep{Teplow_06_AccChemRes}. 

The idea of applying the DMD approach to study protein
folding was proposed in 1996 by Zhou {\it et
al.}~\citep{Zhou_96_PRL}.  Soon after, the method was
combined with a one-bead protein model to study folding of a
model three-helix bundle
protein~\citep{Dokholyan_98_FD,Dokholyan_00_JMB,Zhou_97_PNAS,Zhou_97_JCHP,Zhou_99_JMB}. 
In 2004, Peng {\it et al.} used DMD with two-bead protein
model to study aggregation of an ensemble of 28 A$\beta$40
peptides into a fibrillar structure~\citep{Peng_04_PRE}. 
Smith and Hall introduced four-bead protein model in
combination with the DMD, and showed a cooperative transition
of a polyalanine chain into an $\alpha$-helical conformation
without any {\it a priori} knowledge of the native
state~\citep{Smith_01_PSFG_a,Smith_01_PSFG_b}.  Using the
four-bead protein model and hydrogen bond interactions in
combination with the DMD on a single 16-residue polyalanine
chain, Ding {\it et al.} demonstrated a temperature-induced
conformational change from the $\alpha$-helix to the
$\beta$-hairpin conformation~\citep{Ding_03_PSFG}.  Urbanc
{\it et al.} studied folding and dimer formation using DMD
with the four-bead protein model, and investigated stability
of dimer conformations predicted by DMD approach using
all-atom MD simulations~\citep{Urbanc_04_BJ}.  Lam {\it et
al.} used the same model to study the A$\beta$42 folding and
its temperature dependence~\citep{Lam_06_Conf}.  The results
of Lam {\it et al.} were in a good qualitative agreement with
an all-atom study using implicit
solvent~\citep{Baumketner_06_PS_b} and, importantly, consistent
with the temperature dependence of A$\beta$ secondary
structure, experimentally determined by Gursky and
Aleshkov and Lim {\it et al.}~\citep{Gursky_00_BBA,Lim_07_BBRC}.

Recently, we studied oligomer formation using a four-bead
model with backbone hydrogen bond interactions and the amino
acid-specific hydropathic interactions, but no effective
EIs~\citep{Urbanc_04_PNAS}.  We observed that dimers are the
most abundant among the low molecular weight A$\beta$40
oligomers and that the frequency of trimers and higher-order
oligomers decreases monotonically.  In contrast, the
A$\beta$42 oligomer size distribution was bimodal, with
significantly more pentamers than A$\beta$40.  Multimodal and
unimodal oligomer size distributions are discriminating
properties of A$\beta$42 and A$\beta$40, respectively, as
observed \textit{in vitro} by PICUP~\citep{Bitan_03_PNAS}. 
Experimentally-detected pentamer/hexamer A$\beta$42 oligomers
were termed paranuclei.  Existence of A$\beta$42 paranuclei
and their homotypical assemblies, ``oligo-paranuclei", has
been independently confirmed by a combination of ion mobility
and mass spectrometry~\citep{Bernstein_05_JACS}. 
Importantly, paranucleus-like assemblies have been detected
\textit{in vivo} in the form of dodecameric assemblies termed
ADDLs~\citep{Gong_03_PNAS},
globulomers~\citep{Barghorn_05_JNC}, and A$\beta
^{\star}$56~\citep{Lesne_06_Nature}.  \textit{In vitro}
studies showed that oxidation of M35 blocks A$\beta$42
paranucleus formation~\citep{Bitan_03_JACS}.  A$\beta$
without oxidated M35 displays both
accelerated~\citep{Seilheimer_97_JSB,Snyder_94_BiophysJ} and
delayed~\citep{Watson_98_Biochem} fibrillogenesis rate
relative to wild type A$\beta$.  Analysis of intramolecular
contacts in A$\beta$40 and A$\beta$42 pentamers in our
\textit{in silico} study also showed that M35 forms contacts
with I41 and A42 in A$\beta$42~\citep{Urbanc_04_PNAS},
providing an explanation of the above experimental
results~\citep{Bitan_03_JACS}.  In addition, our prior study
indicated that A$\beta$42 monomers but not A$\beta$40
monomers are characterized by a turn structure, centered at
G37-G38, and that this turn structure was more prominent in
large oligomers~\citep{Urbanc_04_PNAS}.  This result is
consistent with recent proteolysis results using A$\beta$40
and A$\beta$42~\citep{Lazo_05_Pro_Sci}.

There is indirect {\it in vitro} as well as {\it in silico}
evidence suggesting that EIs play a significant role in both
A$\beta$
folding~\citep{Bitan_03_JBC,Lazo_05_Pro_Sci,Borreguero_05_PNAS,Cruz_05_PNAS}
and A$\beta$ fibril
formation~\citep{Petkova_06_Biochemistry,Petkova_02_PNAS,Sciarretta_05_Biochem}. 
In the present study, we follow the protocols of our previous
study~\citep{Urbanc_04_PNAS} using DMD and four-bead protein
model with amino acid-specific
interactions~\citep{Urbanc_06_ME} to elucidate the role of
EIs between pairs of charged amino acids (D, E, K, and R) on
folding and oligomerization of A$\beta$40 and A$\beta$42.

\section*{Methods}

For our simulation method, we use DMD
simulations~\citep{Rapaport_97_Book}.  The main
simplification in this method is to replace continuous
interparticle potentials by a square-well or a combination of
square-well potentials.  As a result, particles move along
straight lines with constant velocities until a pair of
particles reaches a distance at which the interparticle
potential is discontinuous.  A collision event then takes
place during which the velocities and directions of the
particles are updated while preserving the total kinetic
energy, momenta, and angular momenta.  Because DMD is
event-driven, it is faster than all-atom MD.  Our DMD
approach using coarse-grained protein models has been
described in detail elsewhere~\citep{Urbanc_06_ME}. 

Here we use a four-bead protein model with backbone hydrogen
bonding, effective hydropathic interactions and EIs.  We use
the four-bead model with hydrogen bonding, introduced by Ding
{\it et al.}~\citep{Ding_03_PSFG}, then further generalized
by Urbanc {\it et al.}~\citep{Urbanc_04_PNAS} to include
amino acid-specific hydropathic and electrostatic
interactions.  In the four-bead model, the backbone is
represented by three beads, corresponding to the amide ($N$),
the $\alpha$-carbon ($C_\alpha$), and the carbonyl
($C^{\prime}$) groups.  Each side-chain is represented by one
bead ($C_\beta$).  G, which lacks a side-chain, has no
$C_\beta$ bead.  As the carbonyl oxygen and the amide
hydrogen are not explicitly present, an effective backbone
hydrogen bond is introduced between the nitrogen atom $N_i$
of the $i-th$ amino acid and the carbon atom $C_j$ of the
$j-th$ amino acid.  Because the solvent is not explicitly
present in our DMD approach, effective interactions between
the side-chain atoms are introduced to mimic the solvent
effects.  The relative strength of hydropathic interactions
between pairs of side chain beads is based on the
Kyte-Doolittle hydropathy scale~\citep{Kyte_82_JMB}.  When
two hydrophobic side chain beads are within the interaction
range of $0.75$~nm, they interact through a one-step
attractive potential.  When two hydrophilic side chain beads
are within the same interaction distance, they interact
through a one-step repulsive potential.  In our model, the
hydrophobic amino acids are A, C, F, L, M, I, and V.  The
hydrophilic amino acids are D, E, H, K, N, Q, and R.  The
side chains of the remaining amino acids G, P, S, T, W, and Y
interact only through a hard-core repulsion.  The EIs are
implemented by assigning a two-step potential with two
interaction distances, $0.60$~nm and $0.75$~nm, as described
elsewhere~\citep{Urbanc_06_ME}.  When two beads with the same
charge are at the interaction distance, they interact through
a positive (repulsive) two-step potential.  Two oppositely
charged beads interact through a negative (attractive)
two-step potential. 

We set the potential energy of the hydrogen bond, $E_{HB}$,
which in proteins is typically in the range $1-5$
kcal/mol~\citep{Sheu_03_PNAS}, to unit energy ($E_{HB} = 1$). 
We set the potential energy of the hydropathic interactions
$E_{HP} = 0.3$.  Experimental free energy of salt bridge
formation is estimated to be in the range $0.7-1.7$
kcal/mol~\citep{Luisi_03_Biochem}, thus we choose the
potential energy of EIs, $E_{CH} = 0.6$.  Using the unit of
temperature $E_{HB}/k_{B}$ where $k_{B}$ is Boltzmann's
constant, we estimate that $T=0.15$ is appropriate for
simulating physiological temperatures.  We perform DMD
simulations in the canonical ensemble (NVT) using the
Berendsen thermostat algorithm~\citep{Berendsen_84_JCHP}.

Because we treat the solvent in our DMD approach implicitly,
the effective interactions between the side-chain beads
include not only protein-protein but also protein-solvent
interactions.  Thus, there are no generic interaction
parameters that would be independent of the environment. 
Moreover, because different proteins may interact with the
solvent in different ways, the implicit effect of the solvent
and thus the interaction parameters may depend on the
particular protein sequence.  The complexity of
protein-protein and protein-solvent interactions represents a
challenge in protein structure prediction where even the most
successful specialized models fail on certain
targets~\citep{Bradley_05_PROTEINS}.  The question of how
general is a particular choice of interaction parameters in
our DMD approach is a topics of future studies. 

\section*{Results and Discussion}

We apply the four-bead model with hydrogen bonding and amino
acid-specific interactions due to hydropathy and charge and
use DMD with implicit solvent to study A$\beta$40 and
A$\beta$42 oligomer formation.  Due to simplifications in
protein description and implicit solvent, our DMD approach is
efficient enough to allow for a study of the whole process
starting from unfolded separated peptides to formation of
quasi-stable A$\beta$ oligomers with well-defined size
distributions.  In our protein model, each side chain is
replaced by at most one bead, a significant simplification
considering side-chain diversity.  However, recent developments in understanding of protein folding and assembly
show that despite the complexity of the process as a whole, the underlying fundamental physics is
simple~\citep{Baker_00_NATURE,Dobson_04_METHODS}.  It is
believed that the patterns of hydrophobic and hydrophilic
residues, rather than the highly specific characters of the
individual residues involved, play an important
role~\citep{Bowie_91_SCIENCE,Finkelstein_91_NATURE}.  This is
consistent with our prior simulation results where we showed
that amino acid-specific interactions due to hydropathy
itself are sufficient~\citep{Urbanc_04_PNAS} for accounting for
the experimentally observed~\cite{Bitan_03_PNAS} oligomer
size distribution differences between A$\beta$40 and
A$\beta$42.  Here, we apply the same model, with the addition
of Coulombic interactions between pairs of charged amino
acids, to study the effect of EIs on A$\beta$40 and
A$\beta$42 oligomer formation.

The primary structure of A$\beta$42 is DAEFRHDSGYEVHHQKLVFFAEDVG\newline
SNKGAIIGLMVGGVVIA.  The primary
structure of A$\beta$40 is identical, except that the last two
amino acids, I and A, are missing.  We define the following
peptide regions:  ({\it i}) the N-terminal region D1-K16
(NTR); ({\it ii}) the central hydrophobic cluster L17-A21
(CHC); ({\it iii}) the turn A region E22-G29 (TRA); ({\it
iv}) the mid-hydrophobic region A30-M35 (MHR); ({\it v}) the
turn B region V36-V39 (TRB); and ({\it vi}) the C-terminal
region V40/V40-A42 (CTR).  The CTR of A$\beta$40 consists of
only one amino acid, V40.

We simulate eight oligomerization trajectories for A$\beta$40
and A$\beta$42 each, starting from spatially separated
peptides.  Each initial configuration consists of 32
A$\beta$40 (A$\beta$42) peptides with a zero potential energy
and with randomized spatial positions and randomized initial
velocities of atoms within a cubic box of side $25$~nm.  The
molar concentration is $\sim 3.4$~mM.  This initial setup
follows the protocol of our prior
publication~\citep{Urbanc_04_PNAS}.  The concentration in our
simulation is $10-100$ times higher than that studied
experimentally~\citep{Bitan_03_PNAS}.  Lowering the
concentration is possible only at a high cost of efficiency
of our approach.  As shown in a recent study by Nguyen and
Hall~\citep{Nguyen_06_JACS}, lowering the concentration may
give rise to $\alpha$-helical aggregates at low temperatures,
possibly altering the assembly pathways, a problem to be
addressed in future studies.

The energy is in our approach normalized to the potential
energy of the hydrogen bond $E_{HB} = 1$.  Temperature is
expressed in units of energy and also normalized to $E_{HB}$. 
The maximal potential energy of the hydrophobic/hydrophilic
interaction is set to $E_{CH} = 0.6/E_{HB} = 0.6$.  The
N-terminal amine group and the C-terminal carboxyl group are
non-charged.

Hydrogen bonding is the same for all amino acids and
represents the basic interaction needed to model the
secondary structure, $\alpha$-helix and $\beta$-strand,
formation.  When only the hydrogen bond interactions are
allowed ($E_{HB} = 1, E_{HP} = 0,\ and\ E_{CH} = 0$), a single
planar $\beta$-sheet aggregate is
formed~\citep{Urbanc_04_BJ,Urbanc_06_ME}.  Thus, only
hydrogen bond interaction is not enough for description of
spherical oligomers with only small amounts of secondary
structure.  Recently, we introduced the effective
hydrophobic/hydrophilic interactions which are amino
acid-specific to mimic the effect of aqueous
solution~\citep{Urbanc_04_PNAS}.  Using the hydrogen bonding
and effective hydropathic interactions but no EIs ($E_{HB} =
1, E_{HP} = 0.3,\ and\ E_{CH} = 0$), we found spherical
A$\beta$ aggregates with a dense hydrophobic core and with
the hydrophilic N-termini comprising the
surface~\citep{Urbanc_04_PNAS}. 

\begin{figure}[ht]
\centering
\subfigure[][]{\label{conformation_no_EI_Ab42_monomer}
\includegraphics*[width=3.8cm]{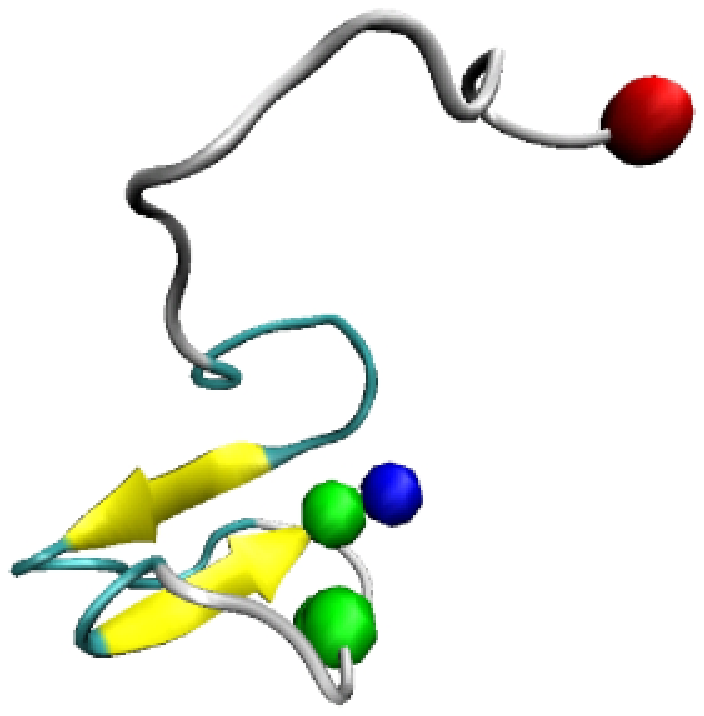}}
\hspace{0.8cm}
\subfigure[][]{\label{conformation_EI_Ab42_monomer}
\includegraphics*[width=3.8cm]{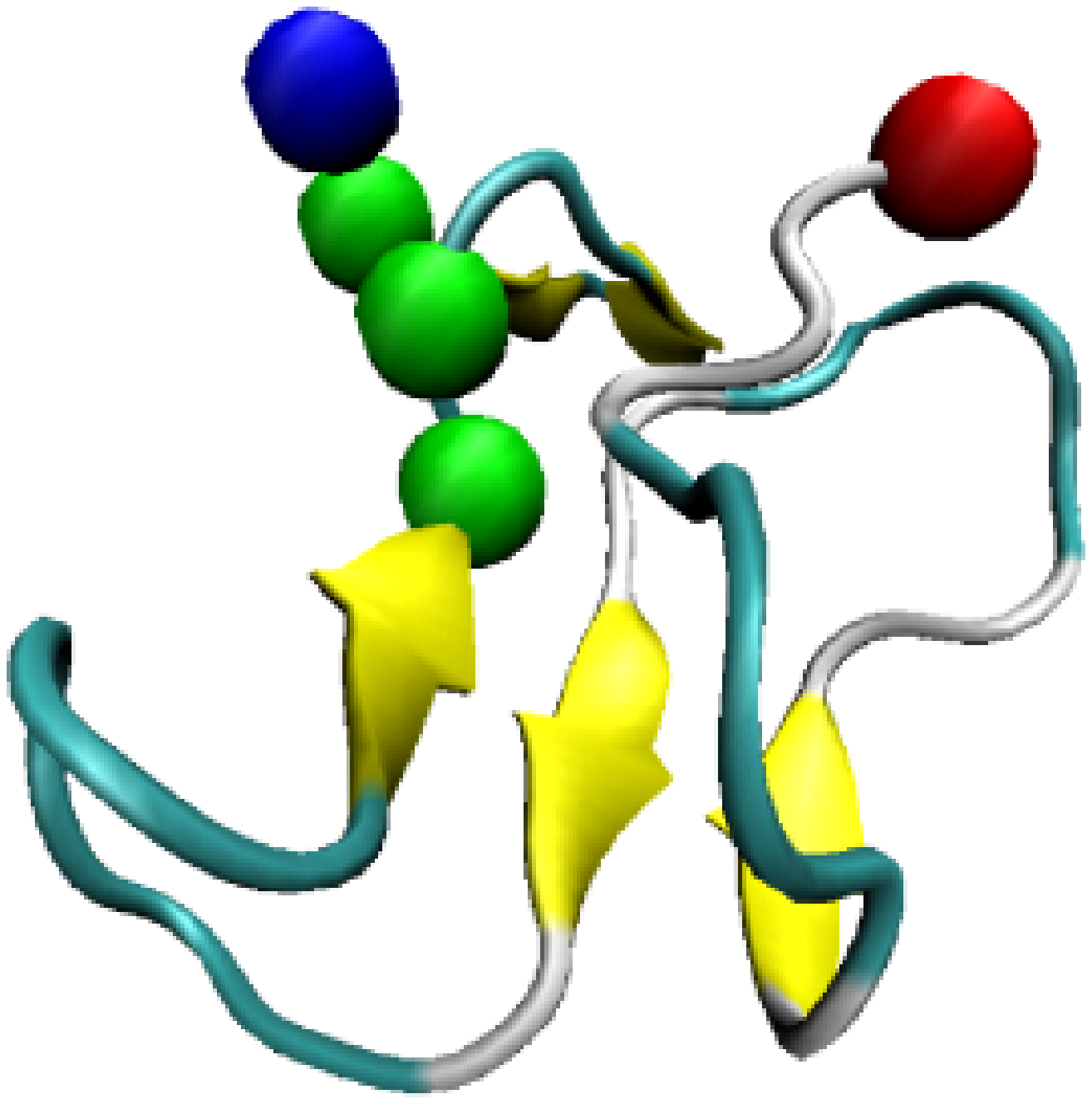}}

\vspace{0.8cm}

\subfigure[][]{\label{conformation_no_EI_Ab42_dimer}
\includegraphics*[width=3.8cm]{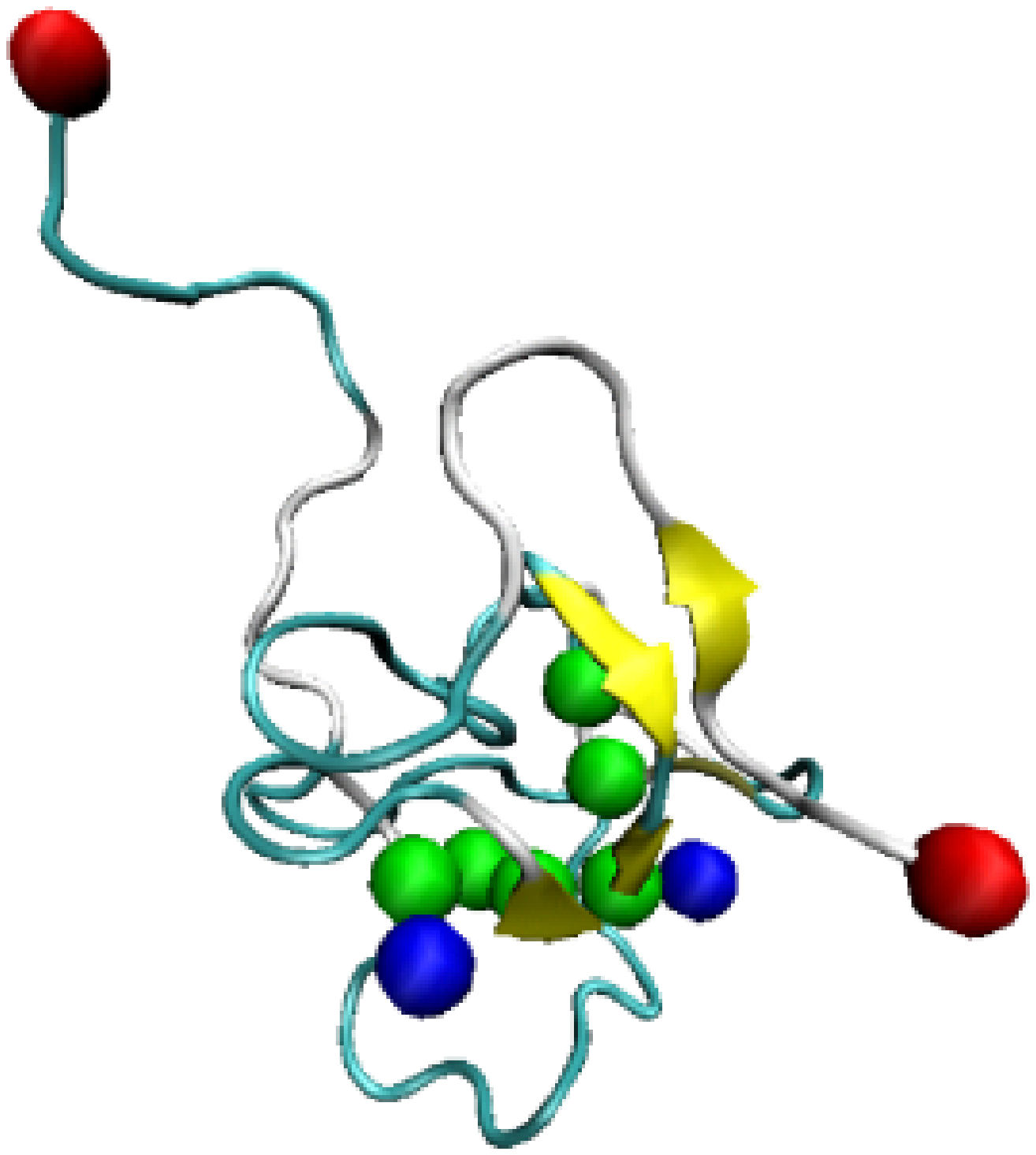}}
\hspace{0.8cm}
\subfigure[][]{\label{conformation_EI_Ab42_dimer}
\includegraphics*[width=3.8cm]{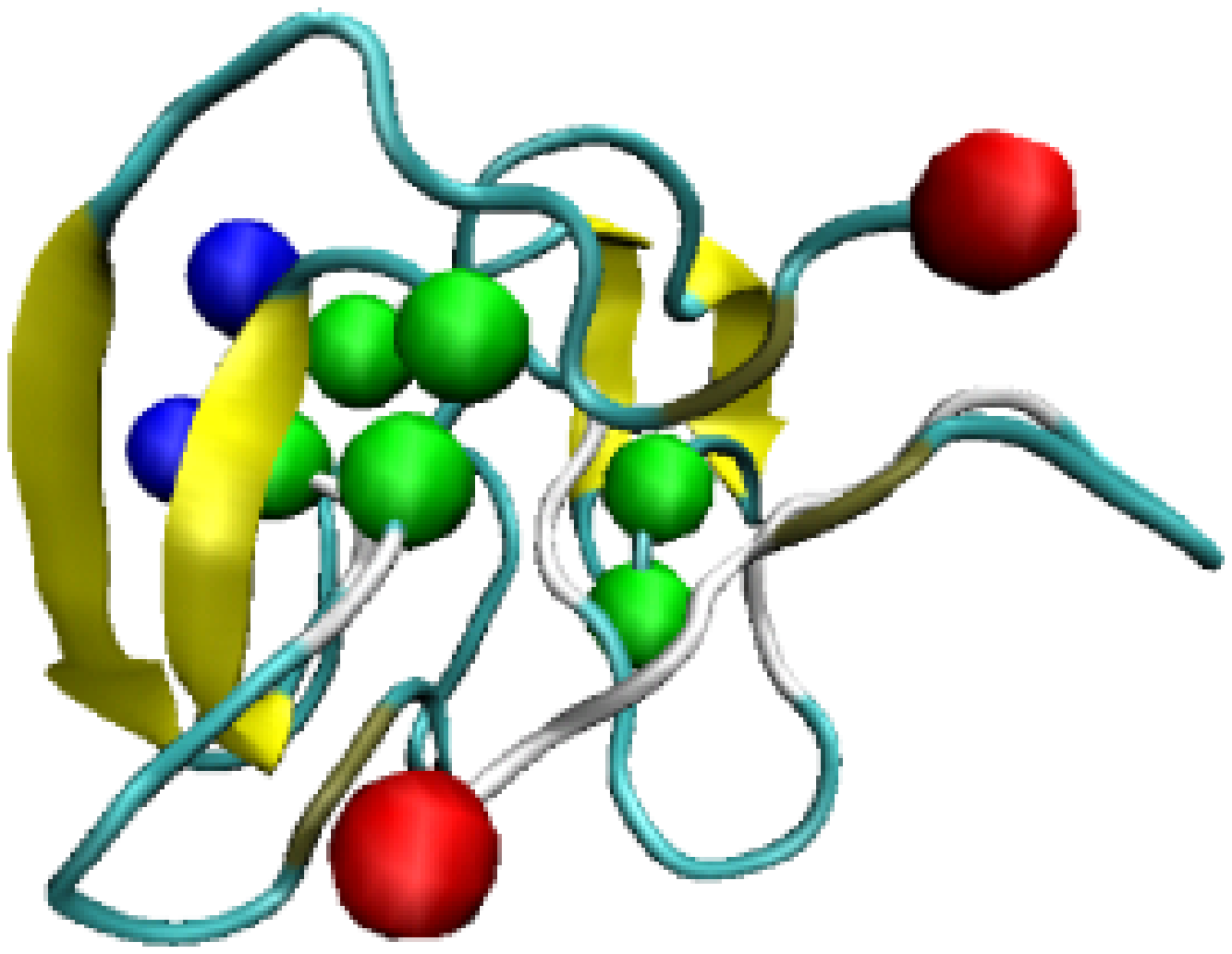}}

\vspace{0.8cm}

\subfigure[][]{\label{conformation_no_EI_Ab42_pentamer}
\includegraphics*[width=3.8cm]{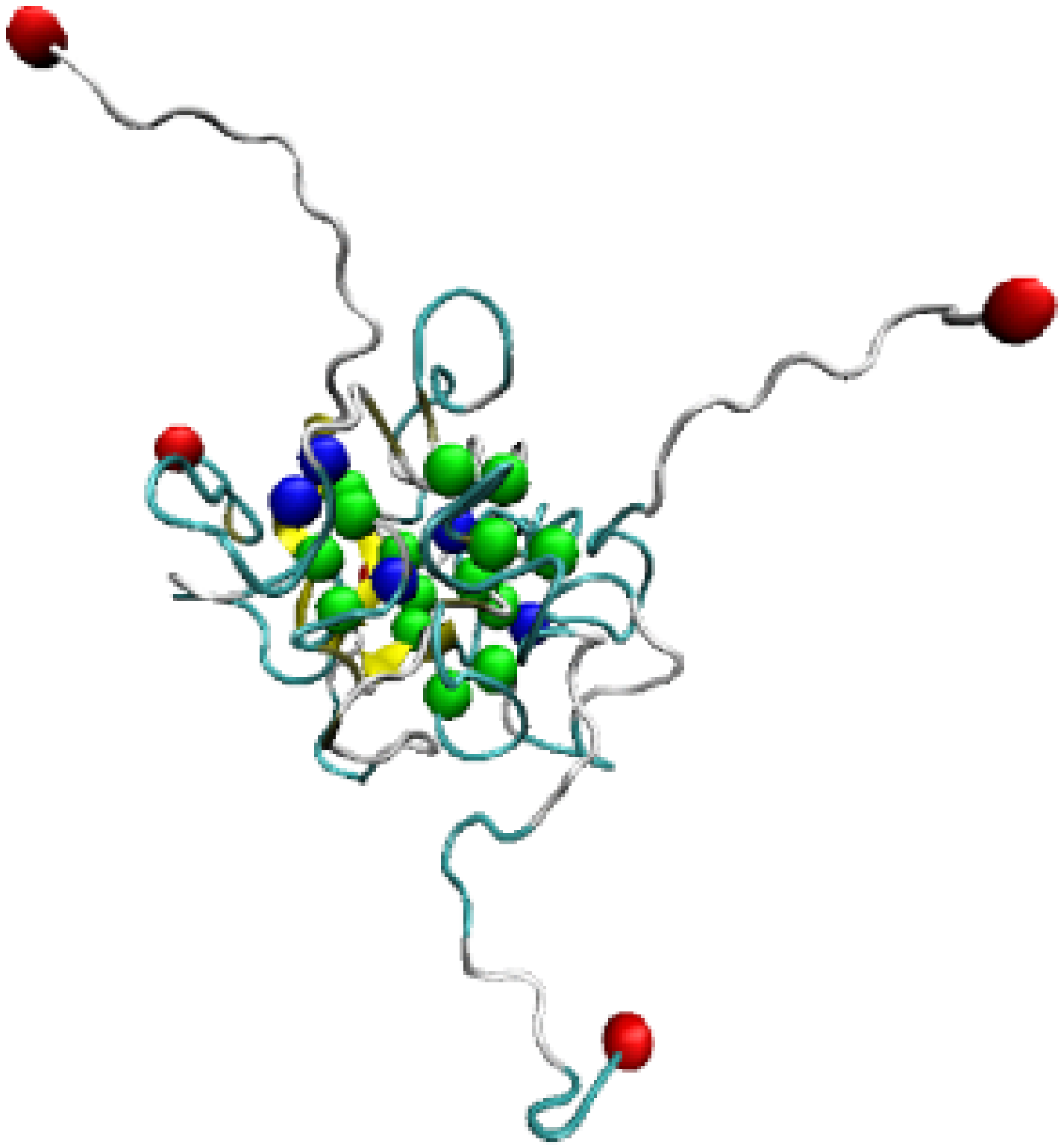}}
\hspace{0.8cm}
\subfigure[][]{\label{conformation_EI_Ab42_pentamer}
\includegraphics*[width=3.8cm]{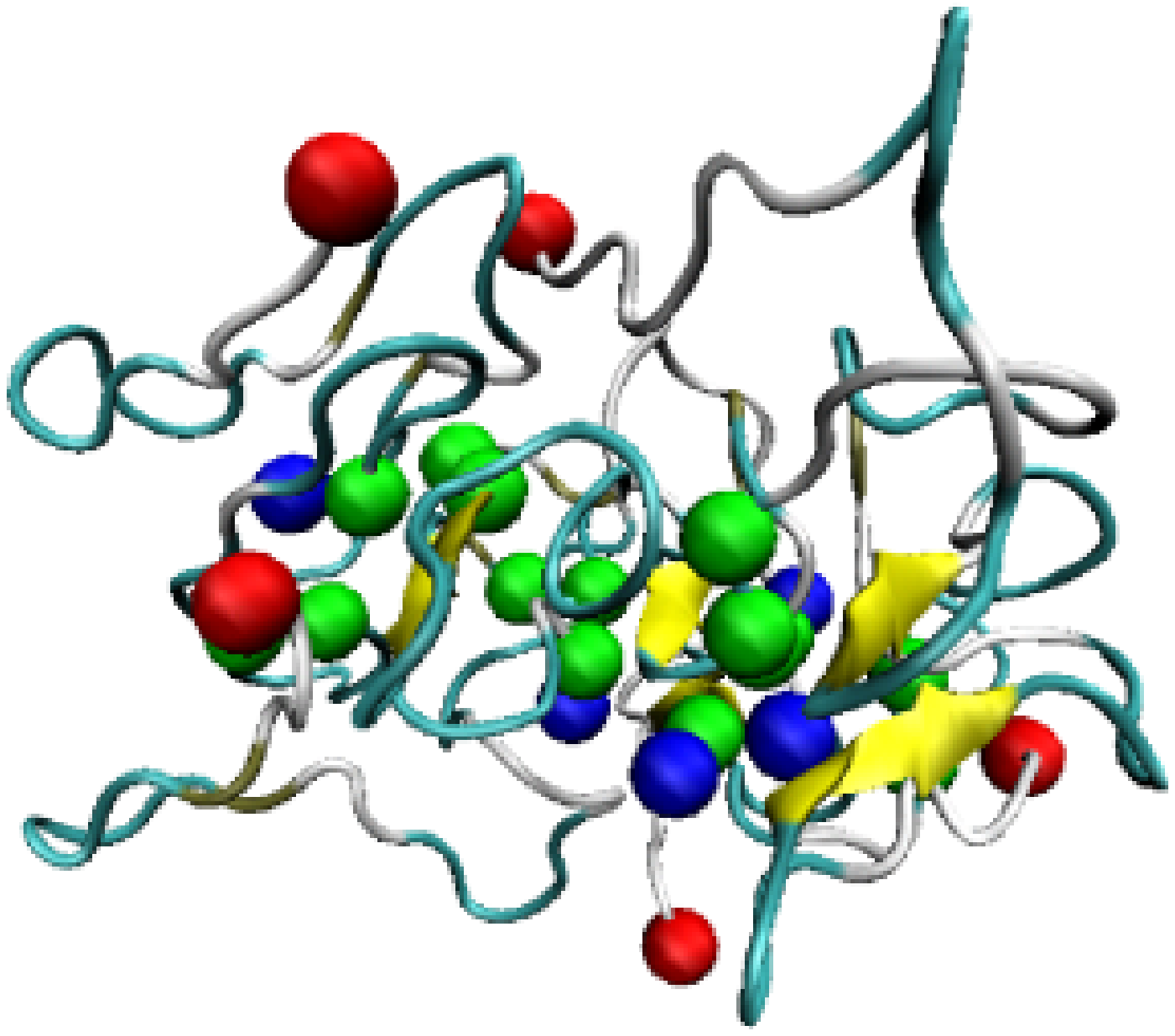}}
\caption{}
\label{conformations}
 
\end{figure}

The aim of the present study is to explore the effects of EIs
on oligomer formation of A$\beta$40 and A$\beta$42.  The
question of how EIs affect the aggregation is intriguing
because most of the charged amino acids are at the N-part of
the molecule:  six of nine charged amino acids are within the
D1-K16 fragment as opposed to the hydrophobic residues which
are concentrated in the remaining fragment L17-V40/A42. 
Fig.~\ref{conformations} shows typical conformations of a
folded monomer, dimer, and pentamer of A$\beta$42 in the
absence and presence of EIs.  Similar conformations are found
in the case of A$\beta$40 (data not shown).  We observe
various topologies at a fixed oligomer size, which is
consistent with findings by Huet {\it et
al.}~\citep{Huet_06_BJ}.  To gain more quantitative insight
into the oligomer formation and structure, we quantify the
oligomer size distributions, calculate the intra- and
intermolecular contact maps, secondary structure
propensities, and Ramachandran plots for each A$\beta$40 and
A$\beta$42 alloform separately.

\subsection*{A$\beta$40 and A$\beta$42 oligomer size
distributions}

All simulations are 10 million simulation steps long. Initially,
all the oligomer size distributions are peaked at monomers and
the oligomer size distributions of A$\beta$40 and A$\beta$42 
are equivalent. The difference between A$\beta$40 and A$\beta$42
size distributions develops steadily with increasing simulation
time and at $\sim$~6 million steps the difference between 
A$\beta$40 and A$\beta$42 oligomer size distributions becomes
statistically significant as determined by applying the
$\chi^{2}$-test~\citep{NumericalRecipes_92}. When comparing
oligomer size distributions of each alloform separately 
at 8.0, 8.5, 9.0, 9.5, and 10.0 million steps, we find
that within this time window size distributions do not
differ significantly. However, the number of monomers and
oligomers of all sizes is variable. Each of the final
oligomer size distributions is obtained by first average
over all 8 trajectories at a fixed simulation time, and then
the resulting ensemble averages are averaged over the 
simulation times of  8.0, 8.5, 9.0, 9.5, and 10.0 million steps.

We have shown previously that A$\beta$40 and A$\beta$42
oligomer size distributions in the absence of EIs ($E_{CH} =
0$) are significantly different
(Fig.~\ref{oligomer_size_no_EI_vs_EI}(a))~\citep{Urbanc_04_PNAS}. 
A$\beta$40 and A$\beta$42 oligomer size distributions in the
presence of EIs ($E_{CH} = 0.6$) are significantly shifted
towards larger oligomers, as shown in
Fig.~\ref{oligomer_size_no_EI_vs_EI}(b).  Comparing the
A$\beta$40 and A$\beta$42 oligomer size distributions by
applying the $\chi^{2}$-test, we conclude that in the
presence of EIs, the distributions are significantly
different ($p < 0.01$).

In the presence of EIs, the average size of A$\beta$40
oligomers increases from 3.0 to 5.2 molecules, and the
average size of A$\beta$42 oligomers increases from 3.7 to
6.2 molecules.  These results suggest that EIs facilitate
aggregation.  A$\beta$42 forms significantly more nonamers
and larger oligomers compared to A$\beta$40.  The A$\beta$40
size distribution is unimodal with a peak at tetramers.  The
A$\beta$42 distribution contains a trimer peak and two
additional peaks, at $n=9$ (nonamer) and $n=14$
(tetradecamer), neither of which is present in the A$\beta$40
distribution.  A multimodal oligomer size distribution was
observed experimentally with A$\beta$42, but not with
A$\beta$40~\citep{Bitan_03_PNAS}. 

In our simulations, the N- and C-termini are uncharged,
whereas in the experimental studies, the N-terminus is
positively charged ($NH_3^+$) and the C-terminus is
negatively charged
($COO-$)~\citep{Bitan_03_PNAS,Bitan_03_JBC}.  Observation of
high-order oligomers in our simulations is consistent with
{\it in vitro} results in which the C-terminal carboxyl group
was replaced by the electrostatically neutral carboxamide,
resulting in a greater abundance of high molecular weight
oligomers~\citep{Bitan_03_JBC}.  Our simulation results, in
combination with experimental findings, thus suggest that
inclusion of charged termini, in particular the C-terminal
negative charge, will moderate formation of A$\beta$
oligomers.  This hypothesis will be tested in future
computational and experimental studies.

\subsection*{Secondary structure of A$\beta$ monomers}

We calculate the secondary structure propensities on each
folded monomer separately using the STRIDE
program~\citep{Heinig_04_NAR} and then average over different
conformations to obtain the average values of the
$\alpha$-helix, turn, and $\beta$-strand propensities per
amino acid.  At 1 million (M) step, the potential energy of individual
monomers is stabilized (data not shown), thus we consider
monomers to be in a folded state at 1M step.

Folded monomers do not have a significant amount of
$\alpha$-helix structure (data not shown). 
Figs.~\ref{monomer_secondary_structure} (a) and (b) show the
turn propensity per amino acid for folded A$\beta$40 and
A$\beta$42 monomers in the absence and presence of EIs.  A
dramatic effect of EIs on the turn propensities in both
alloforms is observed in the region A21-A30.  In the absence
of EIs this region is characterized by two turns, the first
at A21-V24 and the second at S26-G29.  In the presence of
EIs, only a single turn within the region V24-G29 remains. 

Figs.~\ref{monomer_secondary_structure} (c) and (d) show the
$\beta$-strand propensity per amino acid for folded
A$\beta$40 and A$\beta$42 monomers in the absence and
presence of EIs.  As a result of EIs in both alloforms, the
regions A21-D23 and K28-I31 show an increased $\beta$-strand
propensity.  In A$\beta$40 monomers the regions A2-F4 and
L34-G38 show a decreased $\beta$-strand propensity due to
EIs.  In A$\beta$42 monomers the regions R5-H6 and L34-V39
show a slightly decreased $\beta$-strand propensity due to
EIs.  Notice that the $\beta$-strand propensity per amino
acid is below 40\% for A$\beta$40 and below 30\% for
A$\beta$42.  The number of turns and consequently also the
number of $\beta$-strand regions in the A$\beta$42 monomer
(5) is bigger than in the A$\beta$40 monomer (4), indicating
a more compact structure of the A$\beta$42 monomer as
compared to the A$\beta$40 monomer, a consequence of a
strongly hydrophobic CTR in A$\beta$42, which introduces an
additional turn centered at G37-G38.  The average turn and
$\beta$-strand contents of A$\beta$40 and A$\beta$42 folded
monomers are displayed in Table I.  These contents are
calculated from propensities per residue by averaging over
all residues in the the peptide.  Table I shows that for both
A$\beta$40 and A$\beta$42 the average turn content is in the
range 43-45\% while the average $\beta$-strand content is in
the range 10-12\%.  Neither the average turn nor the average
$\beta$-strand content is strongly affected by EIs. 
 
The above results suggest that even in the presence of EIs,
the A$\beta$ monomer is a collapsed coil with several turns
and some $\beta$-strand but no $\alpha$-helical structure,
which is in agreement with existing experimental
studies~\citep{Gursky_00_BBA,Zhang_00_JSB,Lazo_05_Pro_Sci}. 
The $\beta$-strand propensity of A$\beta$40 monomer as shown
in Fig.~\ref{monomer_strand_propensity_Ab40} is also
consistent with a recent study of A$\beta$40 folding using a
scanning tunnelling microscopy that showed monomers folded
into 3 or 4 domains with some $\beta$-strand
structure~\citep{Losic_06_JSB}.

\subsection*{Intramolecular contacts of folded A$\beta$
monomers}

Here we discuss the effect of EIs on the intramolecular
contacts among pairs of amino acids of folded monomers. 
Initially, monomer peptides are in zero-potential energy
(unfolded) conformations.  At 0.1M steps, over 60\% of
peptides (65.9\% for A$\beta$40, 60.5\% for A$\beta$42) are
folded.  We describe the regions of the most important
contacts between pairs of amino acids.  We first
describe ``short-range" contacts formed within the regions
TRB, MHR, and TRA.  Then, we describe the ``long-range"
contacts between the regions CHC-CTR, CHC-MHR, and CHC-NTR. 

Previous results~\citep{Urbanc_04_PNAS}
showed that while A$\beta$40 and A$\beta$42 monomers both
display strong contacts within the TRA region, strong contact
in the TRB region with a turn centered at G37-G38 are
characteristic of A$\beta$42 only.  This {\it in silico}
difference between A$\beta$40 and A$\beta$42 folding is
consistent with experimental findings by Lazo {\it et
al.}~\citep{Lazo_05_Pro_Sci}.  

In Fig.~\ref{monomer_cm}, we compare
the intramolecular contact maps of A$\beta$40 and
A$\beta$42 in the presence and absence of EIs. 
Fig.~\ref{monomer_cm} shows the region containing the
strongest contact V36-V39 as reported in our previous study
(rectangle~1 in (a) and (c))~\citep{Urbanc_04_PNAS}.  In A$\beta$40
the contacts between the amino acid regions L34-V36 and
V39-V40 are significantly weaker than similar contacts
between L34-V36 and V39-A42 in A$\beta$42.  EIs do not affect
contacts in the TRB region (rectangle~1 in (b) and (d)).  This
result suggests that EIs do not alter the contacts that
contribute to differences between A$\beta$40 and A$\beta$42
folding in the CTR.

A few important contacts in both alloforms in the MHR,
concentrated around the strongest contact I31-L34, bring into
proximity the two MHR regions A30-I32 and L34-V36 and are not
affected by EIs (rectangle~3 in (a)-(d)).  The formation of these
contacts within the MHR is promoted by G33 because glycines
are typically associated with a high turn/loop propensity. 
Contacts between the CTR and MHR are present in both
A$\beta$40 (rectangle~2 in (a)) and A$\beta$42 (rectangle~2 in
(c)), but are significantly stronger in A$\beta$42.  These
contacts are not affected significantly by EIs (rectangle~2
in (b) and (d)).

The central and most abundant contacts in folded monomers of
both alloforms are formed as a consequence of the formation
of the turn involving the TRA region (rectangles~4 and~7 in
(a)-(d)).  The TRA region contains charged amino acids E22,
D23, and K28, thus it is expected that EIs will influence the
contacts in this region.  A strong contact A21-V24 in the TRA
region becomes weaker as a result of EIs (rectangle~7 in
(a)-(d)), which is consistent with the effect of EIs on the
turn propensity in this region, changing a two-turn region
into a one-turn region.  Formation of contacts within the TRA
brings into proximity the CHC and MHR (rectangles~5 in
(a)-(d)).  In both alloforms in the absence of EIs, the CHC
region makes contacts with the MHR with F19-I31 as the
strongest contact (rectangles~5 in (a) and (c)).  EIs enhance
the contacts within and around the TRA region in both
alloforms, making contacts between the regions L17-D23 and
K28-I32 (rectangles~5 in (b) and (d)) stronger.  This
enhanced feature is a consequence of a salt bridge formation
between the oppositely charged D23 and K28.  The TRA region
was recently hypothesized to represent the nucleation region
of A$\beta$ folding~\citep{Lazo_05_Pro_Sci}.  This turn has
been shown to be important in the fibril
structure~\citep{Petkova_06_Biochemistry,Petkova_02_PNAS},
suggesting that this region maintains conformational
stability throughout the folding and assembly of A$\beta$. 
Our results are consistent with this hypothesis as they show
that formation of contacts within the TRA region induces
prominent contacts between the CHC and MHR, resulting in the
highest concentration of intramolecular contacts, involving
the TRA, CHC, and MHR.

In the absence of EIs, the MHR region A30-M35 makes contacts
with both the CHC (rectangle~5 in (a)-(d)) and CTR (rectangle~2
in (a)-(d)).  These contacts do not change significantly in the
presence of EIs.  The difference between A$\beta$40 and
A$\beta$42 is that in A$\beta$40 contacts between the regions
A30-I32 and L34-V36 are stronger than the contacts between
A30-I35 and V39-V40, while in A$\beta$42 the contacts between
the regions A30-I35 and V39-A42 are dominant.  This result
suggests that in A$\beta$42 folding the CTR plays a prominent
role, while in A$\beta$40 the contacts within the MHR and
between MHR and CHC regions are more important.

The contacts between the K16-F19 and E11-H14 become more
pronounced in the presence of EIs due to the EI between the
negatively charged E11 and positively charged K16
(rectangle~8 in (a)-(d)).  A weaker group of contacts within
the NTR between F4-H6 and Y10-V12 is a result of a turn
centered at D7-G9 and hydrophobic attraction F4-V12.  These
contacts are very weak in the absence of EIs (rectangle~9 in
(a) and (c)) but become stronger in the presence of EIs due
to salt bridge R5-E11 (rectangle~9 in (b) and (d)).

Long-range contacts between V39-V40 and CHC are
observed in both A$\beta$40 and A$\beta$42 in the absence of
EIs (rectangle~6 in (a) and (c)).  These contacts remain strong
in the presence of EIs (rectangle~6 in (b) and (d)).  In
A$\beta$42, these contacts are stronger than in A$\beta$40,
both in the absence and presence of EIs.  Another region of
long-range contacts is observed in both alloforms between the
K16-F20 and D1-F4 in the absence of EIs
(rectangle~10 in (a) and (c)).  These contacts become more
pronounced in the presence of EIs due to electrostatic
attraction between the negatively charged D1 and E3 and
positively charged K16 (rectangle~10 in (b) and (d)).  The
long-range contacts between CTR and A2-F4, and MHR and
A2-F4 are also present in both A$\beta$40 and
A$\beta$42 but are weaker than the others and are not
significantly influenced by EIs.

\subsection*{Time progression of A$\beta$ folding events}

Fig.~\ref{monomer_cm_time_evolution} shows time evolution of
A$\beta$40 and A$\beta$42 monomer folding events in the
presence of EIs.  Initially, A$\beta$40 and A$\beta$42
monomers are in zero potential energy, random coil
conformations.  At 1k simulation steps, contacts are formed
between L34-V36 and CTR in both A$\beta$40 and A$\beta$42. 
However, only in A$\beta$42, these contacts are associated
with a turn structure in the TRB region as described in the
previous section.  At 2k steps, the contacts between regions
CHC and TRA, CHC and MHR, CHC and CTR develop in both
A$\beta$40 and A$\beta$42.  These contacts are associated
with a turn structure in the TRA region in both A$\beta$40
and A$\beta$42.  At 4k steps, contacts between NTR and CHC
develop in A$\beta$40.  At 8k steps, as the contacts between
NTR and CHC in A$\beta$40 are more pronounced, these contacts
also emerge in A$\beta$42.  At 0.1M steps, the long-range
contacts between NTR and CTR are formed in both A$\beta$40
and A$\beta$42.  Using the regions defined in
Figs.~\ref{monomer_cm} (b) and (d), the time progression of
contacts follows the numbering 1, 2, 3, ...  10, i.e.,
A$\beta$ folding starts at the C-terminal and progresses
towards the N-terminal.  In A$\beta$40, the turn structure in
the TRA region is the first structural element that is formed
in the process of folding, supporting the hypothesis of Lazo
{\it et al.}~\citep{Lazo_05_Pro_Sci} stating that the region
21-30 nucleates A$\beta$folding.  However, in A$\beta$42 the
turn structure in the TRB region is formed before the
formation of the turn structure in the TRA region.  This
result suggests that in A$\beta$42 the TRB region nucleates
the folding prior to formation of contacts in the TRA region.

\subsection*{Secondary structure of A$\beta$ pentamers and
larger oligomers}

In our previous work~\citep{Urbanc_04_PNAS}, we reported the
secondary structure difference between A$\beta$40 and
A$\beta$42 pentamers that can be found in the NTR and CTR. 
A$\beta$42 pentamers displayed an increased $\beta$-strand
propensity at the V39-I41, while A$\beta$40 pentamers showed
an increased $\beta$-strand propensity at the A2-F4.  Our
present data show that these differences remain intact in the
presence of EIs.

Pentamers and larger oligomers do not have any significant
amount of $\alpha$-helix structure (data not shown). 
Figs.~\ref{pentamer_larger_secondary_structure} (a) and (b)
show the turn propensity per amino acid for A$\beta$40 and
A$\beta$42 pentamers and larger oligomers in the absence and
presence of EIs.  EIs do not affect the turn propensity
significantly.  In A$\beta$42, a slight increase in the turn
propensity due to EIs is found in the region R5-Y10.

Figs.~\ref{pentamer_larger_secondary_structure} (c) and (d)
show the $\beta$-strand propensity per amino acid for
A$\beta$40 and A$\beta$42 pentamers and larger oligomers.  In
both alloforms, the $\beta$-strand propensity in the region
K28-I31 slightly increases and in the region L34-G38
decreases due to EIs.  In the presence of EIs, the
$\beta$-strand propensity in the CHC increases in A$\beta$40,
while it decreases in A$\beta$42 pentamers and larger
oligomers.  

We also calculate the average turn and $\beta$-strand
contents within A$\beta$40 and A$\beta$42 pentamers and
larger oligomers in the absence and presence of EIs.  The
data is shown in Table II.  The average contents are
calculated from propensities per residue by averaging over
all residues in the peptide.  The average turn content is in
the range 41-45\% and the average $\beta$-strand content is
in the range 11-13\%.  There is no significant difference
between the two alloforms and no significant effect due to
EIs.

These results show that pentamers and larger oligomers in our
study have a globular structure dominated by turns and loop
and some $\beta$-strand propensity. EIs change the relative
$\beta$-strand propensities of some regions, but do not
affect significantly the overall secondary structure.

\subsection*{Ramachandran plots of selected amino acids
within the A$\beta$42 pentamers and higher oligomers}

Because our protein model as well as the interactions are
simplified, we tested A$\beta$42 oligomer conformers by
calculating the Ramachandran plots.  We selected the
following 10 amino acids from different regions of the
protein:  D1, Y10, F19, E22, D23, S26, K28, M35, I41, and
A42. 

Our results shown in Fig.~\ref{ramachandran_Ab42} indicate
that both in the absence and presence of EIs, the most
populated ($\Phi$, $\Psi$) region corresponds to the
$\beta$-sheet region.  The exceptions are D1 and A42, the N-
and C-terminal amino acids, due to an increased flexibility
at the two termini, and E22.  Interestingly, E22 shows a
substantially higher propensity to form a right-handed
$alpha$-helix.  Our results show that EIs do not affect these
plots in a significant way.  These results are in qualitative
agreement with A$\beta$ dimer analysis of Huet {\it et al.}
who studied A$\beta$ dimer conformations by all-atom
MD~\citep{Huet_06_BJ}, suggesting that our four-bead model
yields relatively realistic set of $\Phi$ and $\Psi$ angles
and thus adequately accounts for the protein backbone
structure.

\subsection*{Tertiary structure of pentamers and larger
oligomers}

The tertiary structure of A$\beta$ molecules within pentamers
and larger oligomers (Fig.~\ref{cm_pentamer_larger_intra}) is
highly reminiscent of the structure of individual monomers
(compare Figs.~\ref{monomer_cm} and
\ref{cm_pentamer_larger_intra}), suggesting that no major
refolding events are needed in monomers prior to oligomer
formation.  However, there is less involvement of the
N-terminal amino acids and more intramolecular contacts
involving the C-terminal amino acids in A$\beta$ molecules
comprising pentamers and larger oligomers of both alloforms.

There are significant differences between A$\beta$40 and
A$\beta$42 intramolecular contact maps of pentamers and
larger oligomers.  The differences between A$\beta$40 and
A$\beta$42 in the absence of EIs have been described in our
previous work~\citep{Urbanc_04_PNAS} and can be observed
comparing the relative importance of the CHC and CTR:  in
A$\beta$42 the contacts of CTR with MHR and CHC are dominant,
while in A$\beta$40 the CHC plays a dominant role.  In
A$\beta$40 (Figs.~\ref{cm_pentamer_larger_intra}(a) and (b)) the
contacts in regions marked by rectangles~1, 3, 4, and 5 get
weaker due to EIs, while the opposite is true in A$\beta$42
(Figs.~\ref{cm_pentamer_larger_intra}(c) and (d)), where the
contacts within the rectangles~1, 2, 3, 4, and 5 get
stronger.  This effect of EIs on the intramolecular contacts
can only be observed in pentamers and larger oligomers and
not in unassembled monomers.  A$\beta$42 pentamers and
larger oligomers, in the presence of EIs, have significantly
stronger intramolecular contacts than A$\beta$40, suggesting
that A$\beta$42 pentamers and larger oligomers are
intrinsically more stable than their A$\beta$40 counterparts.

Fig.~\ref{ramachandran_Ab42} shows Ramachandran scattering
plot on pentamers and larger oligomers of A$\beta$42.  As
seen from contact map analysis, in the presence of EIs, D1s
are more populated in $\beta$-sheet region, which is the
upper left corner. 

\subsection*{Quaternary structure of pentamers and larger
oligomers}

Intermolecular contact maps indicate contacts among different
A$\beta$ molecules within an oligomer that are most important
in oligomer formation.  Previously, we showed that in
A$\beta$40 pentamers, pairs of the CHC regions show the
highest propensity to interact, whereas in A$\beta$42
pentamers the most frequent contacts are between the CTR of
one peptide and the CHC and MHR of the
other~\citep{Urbanc_04_PNAS}.  That result indicated that the
CTRs are critically involved in aggregation of A$\beta$42 but
not A$\beta$40.

Fig.~\ref{cm_pentamer_larger_inter} shows intermolecular
contact maps of pentamers and larger oligomers of A$\beta$40
and A$\beta$42 in the absence ((a) and (c)) and presence ((b)
and (d)) of EIs.  Perhaps the most surprising overall
observation is that the intermolecular contacts that involve
the CHC, i.e., contacts between pairs of CHCs (rectangle~3 in
(a)-(d)), between the CHC and MHR (rectangle~5 in (a)-(d)),
and between the CHC and CTR (rectangle~6 in (a)-(d)), become
weaker as a consequence of EIs in both alloforms, but this
weakening is more pronounced in A$\beta$40 oligomers.  This
weakening of the contacts involving the CHC due to EIs is
surprising because the CHC is surrounded by charged residues
(K16, E22, and D23).  Thus, we would expect CHCs to interact
pairwise in an anti-parallel fashion to maximize the the
mutual attraction involving hydrophobic residues by
additional salt bridge formation and thus minimize the free
energy.  Instead, our results show that EIs weaken the
contacts between pairs of CHCs.  We also showed that EIs
promote formation of larger oligomers in both A$\beta$40 and
A$\beta$42.  These two results combined imply that weakening
of the contacts between pairs of CHCs in A$\beta$40 oligomers
might actually indirectly promote aggregation into larger
oligomers.

The only exception to the above observation is the region
between D1-R5 and K16-D23 which is rather weak in both
alloforms in the absence of EIs, but gets more pronounced
in particular in A$\beta$42 due to EIs (rectangle~7 in (a)-(d)).

Our results indicate important differences in the way EIs
affect A$\beta$40 and A$\beta$42 oligomers.  In A$\beta$40
oligomers the intermolecular contacts between pairs of CTRs
(rectangle~1 in (a) and (b)), between pairs of MHRs
(rectangle~2 in (a) and (b)), and between the CTR and MHR
(rectangle~4 in (a) and (b)) remain unaffected by EIs.  In
A$\beta$42 oligomers, on the other hand, the intermolecular
contacts in these same regions get stronger even though that
part of A$\beta$42 (MHR and CTR) is free of charge and thus
EIs would not be expected to make a difference.  The
strongest increase in the intermolecular contact intensity in
A$\beta$42 oligomers is between pairs of CTRs (rectangle~1 in
(b) and (d)) and the second strongest is between the CTR and
MHR (rectangle~4 in (b) and (d)).  Thus, in A$\beta$42
oligomers the contacts involving the CHCs get weaker and the
contacts involving the CTRs get stronger due to EIs,
resulting in a significantly larger oligomers.  These results
suggest that in A$\beta$42 the CTRs are most important for
intermolecular assembly into pentamers and larger oligomers. 
The lack of strong intermolecular contacts involving CTRs in
A$\beta$40 suggests that the CTRs are also the main source of
the differences between A$\beta$40 and A$\beta$42 oligomer
formation.  Recently, the importance of the intermolecular
CHC contacts in A$\beta$40 versus the intermolecular CTR
contacts in A$\beta$42 was observed experimentally by Maji {\it et
al.}~\citep{Maji_05_Biochemistry}, in agreement with our present {\it in
silico} results, suggesting the biological relevance of our
DMD approach which is able to capture the essential
differences between A$\beta$40 and A$\beta$42
oligomerization.

\subsection*{Intra and intermolecular hydrogen bonds in pentamers
and larger oligomers}

Here we address the question of how much hydrogen bonds
contribute to intra- and intermolecular contacts in pentamers
and larger oligomers.  We first calculate the probabilities
for forming an intra- or intermolecular hydrogen bond per
amino acid.  The amino acids that are most hydrogen bond
active are shown in Tables III and IV.  Our results show that
even for the amino acids that are most likely to form
hydrogen bonds, probabilities are smaller than 0.20.  The sum
of intra- and intermolecular probabilities per amino acid
does not exceed 0.30/0.40, which is consistent with the
$\beta$-strand propensity per amino acid
(Fig.~\ref{pentamer_larger_secondary_structure}).

Fig.~\ref{cm_hb_intra} shows the intramolecular hydrogen bond
contacts of A$\beta$40 ((a) and (b)) and A$\beta$42 ((c) and
(d)) pentamers and larger oligomers in the absence ((a) and
(c)) and presence ((b) and (d)) of EIs.  These intramolecular
hydrogen bond maps are normalized to the highest value of
intramolecular hydrogen bond formation probability, which is
$<0.09$.  The regions with the highest amount of hydrogen
bonds can be found between the regions K16-V24 and K28-V40. 
In A$\beta$42 oligomers some additional hydrogen bonds are
formed between the MHR and CTR and between the CHC and CTR. 
EIs increase the hydrogen bond probabilities within the TRA
region and between the CHC and MHR due to salt bridge
D23-K28.  This effect is more pronounced in A$\beta$40. 
Interestingly, the strongest intramolecular hydrogen bond
occurs in A$\beta$42 oligomers between F4 and V12, possibly
stabilized by proximity of oppositely charged R5 and E11. 
Why this same hydrogen bond is missing in A$\beta$40
oligomers may be understood by observation that in A$\beta$40
the region A2-F4 forms a $\beta$-strand that is in contact
with the CHC and thus the charged NTR residues (E3 and R5)
are interacting with the charged residues K16 and E22,
preventing R5-E11 from interacting and breaking the F4-V12
hydrogen bond.

The intermolecular hydrogen bonds of A$\beta$40 ((a) and (b))
and A$\beta$42 ((c) and (d)) pentamers and larger oligomers
in the absence ((a) and (c)) and presence ((b) and (d)) of
EIs are presented in Fig.~\ref{cm_hb_inter}.  These
intermolecular contact maps are normalized to the highest
value of intermolecular hydrogen bond probability, which is
$<0.04$.  The probability of intermolecular hydrogen bond
formation is slightly higher in the regions where the
contacts are more pronounced.  EIs do not influence the
intermolecular hydrogen bond formation in any significant
way.

Our results show that the hydrogen bonds present in A$\beta$
pentamers and larger oligomers are not specific,
indicating that oligomers are not characterized by any
particular pattern of hydrogen bonding.  These findings suggest
that hydrogen bonding is mostly a secondary effect occurring
as a consequence of hydrophobic contact formation in the
regions CHC, MHR, and CTR.

\section*{Conclusions}

Because molecular dynamics approach to study proteins using
all-atom representation and explicit solvent is limited to
time scales smaller than $\sim 10^{-6}$~s, we use a
simplified but efficient DMD approach combined with a
four-bead protein model and amino acid-specific interactions
that mimic the effects of a solvent~\citep{Urbanc_06_ME}.  In
our prior work we showed that this approach yields
biologically relevant results, which are consistent with
existing experimental findings on A$\beta$ oligomer formation
and have predictive power allowing for {\it in vitro} and
further {\it in silico} testing~\citep{Urbanc_04_PNAS}.  In
the present work we use the DMD approach to study the effects
of EIs on oligomer formation of A$\beta$40 and A$\beta$42. 
The role of electrostatic interactions, in particular the
salt bridge formation between negatively charged E22/D23 and
positively charged K28 was hypothesized to be important at
early stages of folding as well as at later stages of fibril
formation.  Thus, it is reasonable to expect that EIs may
play an important role at intermediate stages of oligomer
formation.

We analyze the structure of folded A$\beta$40 and A$\beta$42
monomers in the presence and absence of EIs.  We show that
independent of EIs the two alloforms display differences in
folded structure:  in A$\beta$42 there is an additional turn
centered at G37-G38 that is absent in A$\beta$40, leading to
an increased propensity to form $\beta$-strand in the CTR of
only A$\beta$42.  A$\beta$40 monomers also have an additional
$\beta$-strand in the A2-F4 which is not present in
A$\beta$42.  Our results demonstrate that the differences
between the two alloforms are present already at the stage of
folding, prior to assembly.  The existence of a turn
structure centered at G37-G38 is consistent with experimental
findings by Lazo {\it et al.} who showed by using limited
proteolysis that Val39-Val40 in A$\beta$42 but not in A$\beta$40
monomer was protease resistant, indicating that A$\beta$42 but
not A$\beta$40 monomer was structured in the CTR
region~\citep{Lazo_05_Pro_Sci}.  Similar was a conclusion of the
solution NMR study on [Met(O)$^{35}$]A$\beta$40 versus
[Met(O)$^{35}$]A$\beta$42 monomer structure by Riek {\it et al.} 
showing that G29-A42 region is less flexible and thus more
structured in A$\beta$42 than in
A$\beta$40~\citep{Riek_01_EJB}.  By measuring
$^1H_{\alpha}$, $^{13}C_{\alpha}$, and $^{13}C_{\beta}$
chemical shift indices of A$\beta$40 and A$\beta$42 Hou {\it et
al.} recently showed that the C-terminus of A$\beta$42 but
not of A$\beta$40 monomer has a tendency to form
$\beta$-sheet structure~\citep{Hou_04_JACS} which provides
further evidence that our simulation approach yields
biologically relevant results consistent with {\it in vitro}
findings.

Our results indicate that EIs stabilize a turn in the region
D23-K28 by formation of a D23-K28 salt bridge.  A role for
EIs in stabilizing this region has been postulated by Lazo
{\it et al.}~\citep{Lazo_05_Pro_Sci} and further explored
using a more complex united atom DMD
model~\citep{Borreguero_05_PNAS} and all-atom MD in
explicit~\citep{Cruz_05_PNAS} and implicit
solvent~\citep{Baumketner_06_PS_b}.  These studies show that
A$\beta$ folding in the region A21-A30 is driven primarily by
effective hydrophobic attraction between V24 and the butyl
portion of K28, but that EIs help stabilize the region.  In
our model, due to its simplicity, the side chains of V24 and
K28 do not experience attractive interactions.  Despite the
absence of this interaction, we still find this region to be
the most structured in both A$\beta$40 and A$\beta$42
monomers stabilized by D23-K28 salt bridge.  The D23-K28 salt
bridge was suggested to stabilize the A$\beta$40 fibril
structure by Petkova {\it et al.}~\citep{Petkova_02_PNAS}. 
In addition, Sciarretta {\it et al.} have shown an increase
in the rate of A$\beta$40-Lactam (D23/K28) fibrillogenesis by
1000-folds~\citep{Sciarretta_05_Biochem}, providing
additional experimental evidence supporting a critical role
of D23-K28 salt bridge formation.

Comparing the oligomer size distributions of A$\beta$40 and
A$\beta$42 in the presence of EIs with those obtained in the
absence of EIs~\citep{Urbanc_04_PNAS} reveals that EIs
promote formation of larger oligomers while maintaining a
unimodal A$\beta$40 size distribution and a multimodal
A$\beta$42 size distribution, as observed {\it in
vitro}~\citep{Bitan_03_PNAS}.  In our simulations the N- and
C-termini are uncharged in contrast to most experimental
studies with positively charged N- and negatively charged
C-termini.  Our observation that EIs promote formation of
larger oligomers is thus consistent with results of the
experimental study in which the C-terminal carboxyl group was
replaced by the electrostatically neutral carboxamide,
resulting in a greater abundance of high molecular weight
oligomers~\citep{Bitan_03_JBC}.

It is critical to study the structural changes in oligomers
due to EIs and understand which structural changes are
contributing to formation of larger oligomers in both
A$\beta$40 and A$\beta$42.  Our results indicate that in
A$\beta$40 pentamers and larger oligomers, EIs weaken
intramolecular interactions.  In A$\beta$42, in contrast, the
intramolecular contacts in the turn region D23-K28 are
enhanced.  Surprisingly, in both A$\beta$40 and A$\beta$42
oligomers, the intermolecular contacts involving the CHC are
significantly weaker in the presence of EIs.  In addition, in
A$\beta$42 oligomers, the contacts involving the CTR and MHR
get stronger.  These results, combined with the fact that EIs
promote larger oligomers, imply that the intermolecular
interactions between pairs of CHCs in an indirect way oppose
the formation of larger oligomers, while the interactions
between pairs of CTRs, and to a smaller extent also pairs of
MHRs, promote formation of larger oligomers.  Thus,
therapeutic strategies using inhibitors that target the CTR
and MHR may prove successful in either inhibiting formation
of toxic A$\beta$42 oligomers or inducing structural
modifications neutralizing their toxicity.

{\small This work was supported by grants from the National
Institutes of Health (AG023661, NS44147, NS38328, AG18921,
and AG027818), grant A04084 from the American Federation for
Aging Research, grant 2005/2E from the Larry L.  Hillblom
Foundation, an Alzheimer's Association Zenith Fellows award,
and the Petroleum Research Fund.  We are thankful to Stephen
Bechtel, Jr.  for a private donation.}

\newpage
\clearpage


\newpage  
{\bf Figure captions: } 

{Figure 1:  Representative conformations of a monomer, dimer,
and pentamer of A$\beta$42 in the absence ($E_{CH} = 0$) and
presence ($E_{CH} = 0.6$) of EIs.  A monomer conformation (a)
in the absence and (b) presence of EIs.  A dimer conformation
(c) in the absence and (d) presence of EIs.  A pentamer
conformation (e) in the absence and (f) presence of EIs. 
Yellow arrows correspond to the $\beta$-strand structure,
turns are represented by light blue tubes and random
coil-like part are represented by gray tubes.  The N-terminal
D1 is marked as a red sphere, and the C-terminal A42 is
marked as a blue sphere.  I31, I32, and I41, the most
hydrophobic residues, are represented as green spheres.  This
figure is generated by the VMD
package~\citep{Humphrey_96_JMG}.}

{Figure 2:  Oligomer size distributions of A$\beta$40 and
A$\beta$42 at (a) $E_{CH} = 0$ and (b) $E_{CH} = 0.6$.  All
size distributions are averages over the time frames at 8,
8.5, 9, 9.5, and 10 million simulation steps.}

{Figure 3:  The effect of EIs on turn and $\beta$-strand
propensities per residue in folded A$\beta$ monomers in the absence and presence of EIs.  Turn propensities of
(a) A$\beta$40 and (b) A$\beta$42 monomers.  $\beta$-strand
propensities of (c) A$\beta$40 and (d) A$\beta$42 monomers.}

{Figure 4:  Intramolecular contact maps of folded A$\beta$40
and A$\beta$42 monomers at $E_{CH} = 0$ (left column) and
$E_{CH} = 0.6$ (right column).  The strength of the contact
map is color-coded following the rainbow scheme:  from blue
(no contacts) to red (the largest number of contacts).  Each
contact map is an average of over 100 monomer conformations
after 1 million simulation steps.}

{Figure 5:  Detailed time evolution of intramolecular contact
formation during A$\beta$40 (left column) and A$\beta$42
folding (right column).  The strength of the contact map is
color-coded as in Fig.~4.  Each contact map is an average of
over 100 monomer conformations.}

{Figure 6:  The effect of EIs on turn and $\beta$-strand
propensities per residue within A$\beta$ pentamers and larger
oligomers in the absence and presence of EIs.  Turn
propensities of (a) A$\beta$40 and (b) A$\beta$42 pentamers
and larger oligomers.  $\beta$-strand propensities of (c)
A$\beta$40 and (d) A$\beta$42 pentamers and larger
oligomers.}

{Figure 7:  Ramachandran plots of A$\beta$42 pentamers and
larger oligomers for selected residues D1, Y10, F19, E22,
D23, S26, K28, M35, I41, and A42 in the absence and presence
of EIs.  Horizontal and vertical axes correspond to the
angles $\Phi$ and $\Psi$, respectively, both varying from
-180$^\circ$ to 180$^\circ$.  Each plot contains $\sim$640
points corresponding to A$\beta$42 pentamers to decamers
obtained at 8, 8.5, 9, 9.5, and 10 million simulation
steps.  Ramachandran plots are generated using the VMD software package~\citep{Humphrey_96_JMG}.}

{Figure 8:  Intramolecular contact maps of A$\beta$40 and
A$\beta$42 pentamers and larger oligomers at $E_{CH} = 0$ and
$E_{CH} = 0.6$.  Each contact map is an average of over 50
conformations obtained at 8, 8.5, 9, 9.5, and 10 million
simulation steps.}

{Figure 9:  Intermolecular contact maps of A$\beta$40 and
A$\beta$42 pentamers and larger oligomers at $E_{CH} = 0$ and
$E_{CH} = 0.6$.  Each contact map is an average of over 50
conformations at 8, 8.5, 9, 9.5, and 10 million simulation
steps.}

{Figure 10:  Intramolecular hydrogen bond maps of A$\beta$40
and A$\beta$42 pentamers and larger oligomers at $E_{CH} = 0$
and $E_{CH} = 0.6$.  Each map is an average of over 50
conformations at 8, 8.5, 9, 9.5, and 10 million simulation
steps.}

{Figure 11:  Intermolecular hydrogen bond maps of A$\beta$40
and A$\beta$42 pentamers and larger oligomers at $E_{CH} = 0$
and $E_{CH} = 0.6$.  Each map is an average of over 50
conformations at 8, 8.5, 9, 9.5, and 10 million simulation
steps.}

{\bf Table captions: } 

{Table 1:  Average turn and $\beta$-strand propensities per
residue with standard errors within folded A$\beta$40 and
A$\beta$42 monomers. Each value is an average of
over 100 monomer conformations after 1 million simulation steps.}

{Table 2:  Average turn and $\beta$-strand propensities per
residue with standard errors within A$\beta$40 and A$\beta$42
pentamers and larger oligomers.  Each value is an average of
over 50 conformations at 8, 8.5, 9, 9.5, and 10 million
simulation steps.}

{Table 3:  Average hydrogen bond propensities per residue,
showing the five most frequent residues involved in
intramolecular hydrogen bonding within A$\beta$40 and
A$\beta$42 pentamers and larger oligomers.  Each value is an
average of over 50 conformations at 8, 8.5, 9, 9.5, and 10
million simulation steps.}

{Table 4:  Average hydrogen bond propensity per residue,
showing the five most frequent amino acids involved in
intermolecular hydrogen bonding within A$\beta$40 and
A$\beta$42 pentamers and larger oligomers.  Each value is an
average of over 50 conformations at 8, 8.5, 9, 9.5, and 10
million simulation steps.}

\newpage
\clearpage

\newpage
\clearpage
\begin{figure}[ht]
\centering
\subfigure[][]{
\includegraphics*[width=6cm]{fig2a.eps}}
\hspace{1cm}
\subfigure[][]{
\includegraphics*[width=6cm]{fig2b.eps}}  
\caption{}
\label{oligomer_size_no_EI_vs_EI}
 
\end{figure}

\newpage
\clearpage
\begin{figure}[ht]
\centering
\subfigure[][]{\label{monomer_turn_propensity_Ab40}
\includegraphics*[width=6cm]{fig3a.eps}}
\hspace{0.8cm}
\subfigure[][]{\label{monomer_turn_propensity_Ab42}
\includegraphics*[width=6cm]{fig3b.eps}}

\vspace{0.8cm}

\subfigure[][]{\label{monomer_strand_propensity_Ab40}
\includegraphics*[width=6cm]{fig3c.eps}}
\hspace{0.8cm}
\subfigure[][]{\label{monomer_strand_propensity_Ab42}
\includegraphics*[width=6cm]{fig3d.eps}}
\caption{}
\label{monomer_secondary_structure}
 
\end{figure}

\newpage
\clearpage
\begin{figure}[ht]
\begin{center}
$\begin{array}{c}

\includegraphics*[width=14cm]{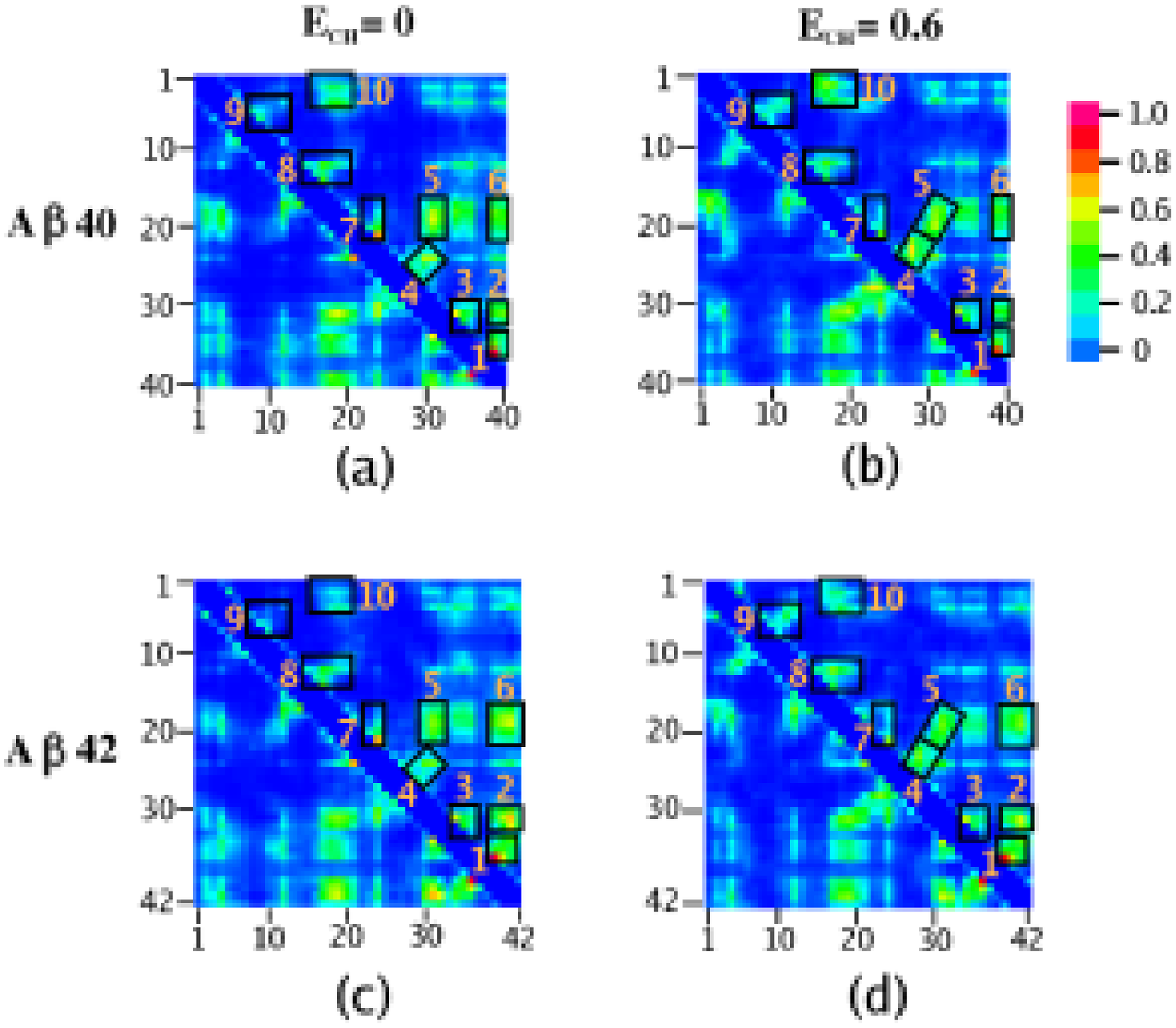}
\\
\end{array}$
\end{center}
\caption{}
\label{monomer_cm}

\end{figure}

\newpage
\clearpage
\begin{figure}[ht]
\centering
\includegraphics*[width=10cm]{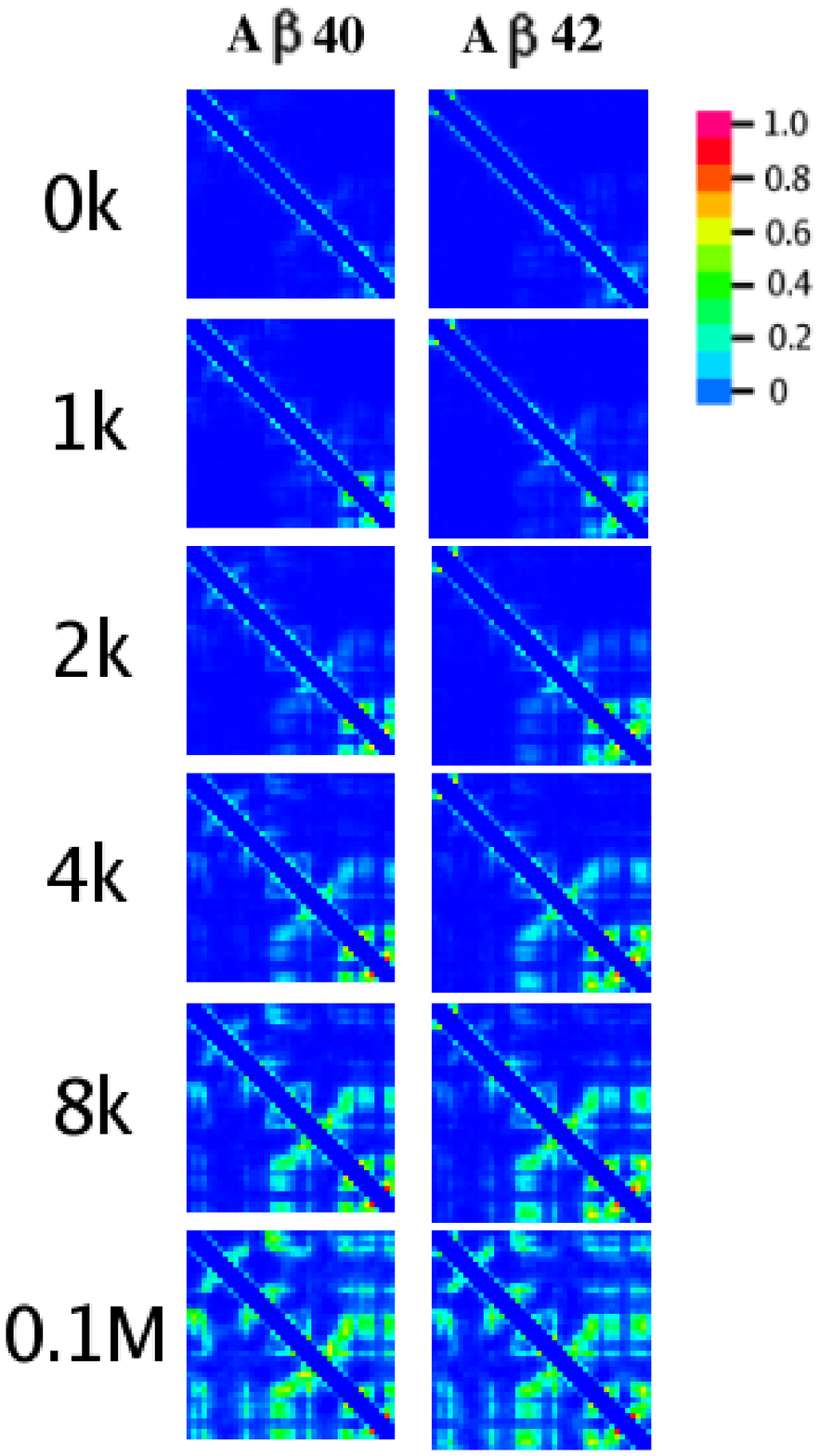}
\caption{}
\label{monomer_cm_time_evolution}
\end{figure}

\newpage
\clearpage
\begin{figure}[ht]
\centering
\subfigure[][]{\label{pentamer_larger_turn_propensity_Ab40}
\includegraphics*[width=6cm]{fig6a.eps}}
\hspace{0.8cm}
\subfigure[][]{\label{pentamer_larger_turn_propensity_Ab42}
\includegraphics*[width=6cm]{fig6b.eps}}

\vspace{0.8cm}

\subfigure[][]{\label{pentamer_larger_strand_propensity_Ab40}
\includegraphics*[width=6cm]{fig6c.eps}}
\hspace{0.8cm}
\subfigure[][]{\label{pentamer_larger_strand_propensity_Ab42}
\includegraphics*[width=6cm]{fig6d.eps}}
\caption{}
\label{pentamer_larger_secondary_structure}
 
\end{figure}

\newpage
\clearpage
\begin{figure}[ht]
\centering
\includegraphics*[width=14cm]{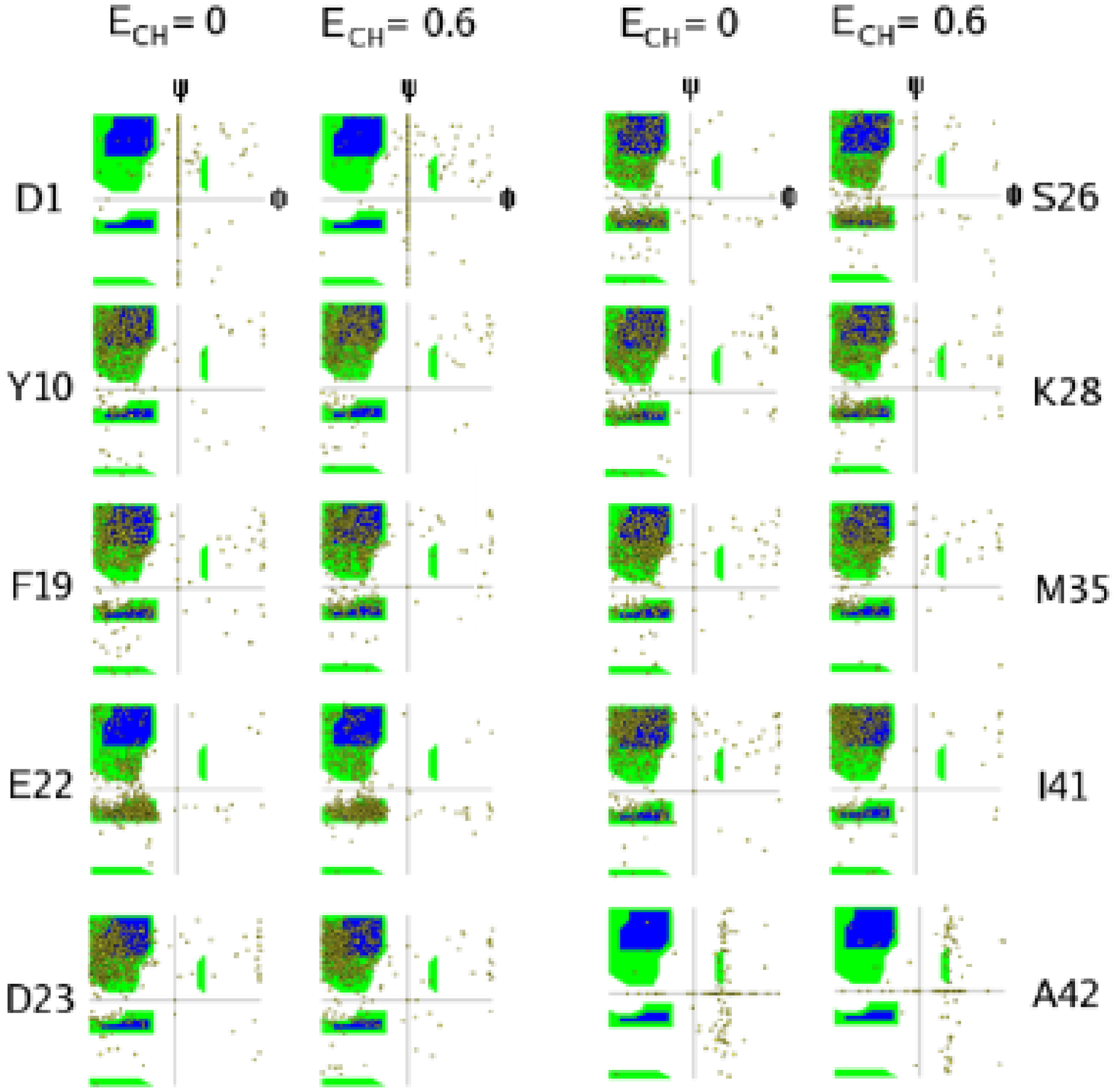}
\caption{}
\label{ramachandran_Ab42}
 
\end{figure}

\newpage
\clearpage
\begin{figure}[ht]
\centering
\includegraphics*[width=14cm]{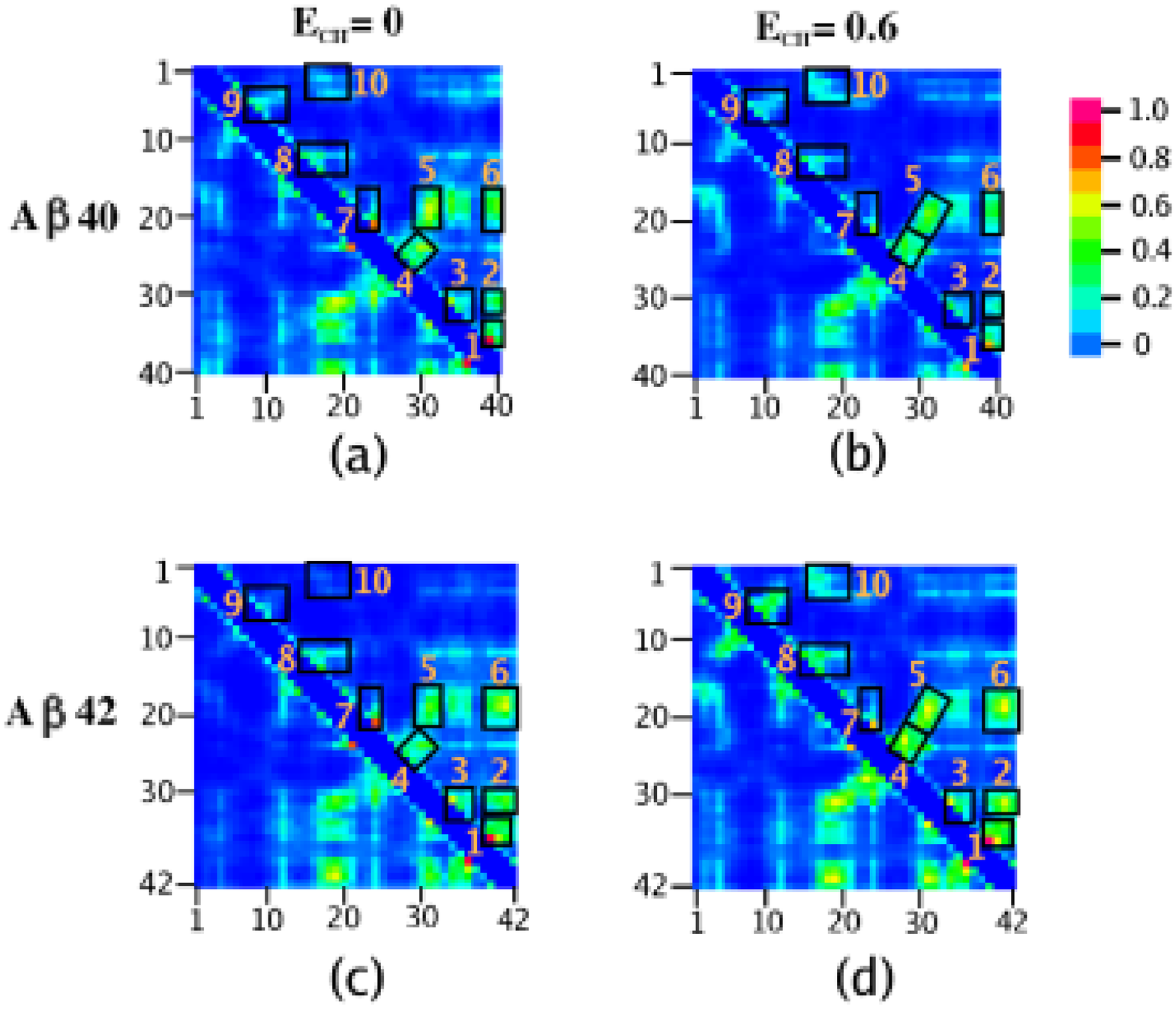}
\caption{}
\label{cm_pentamer_larger_intra}
 
\end{figure}

\newpage
\clearpage
\begin{figure}[ht]
\centering
\includegraphics*[width=14cm]{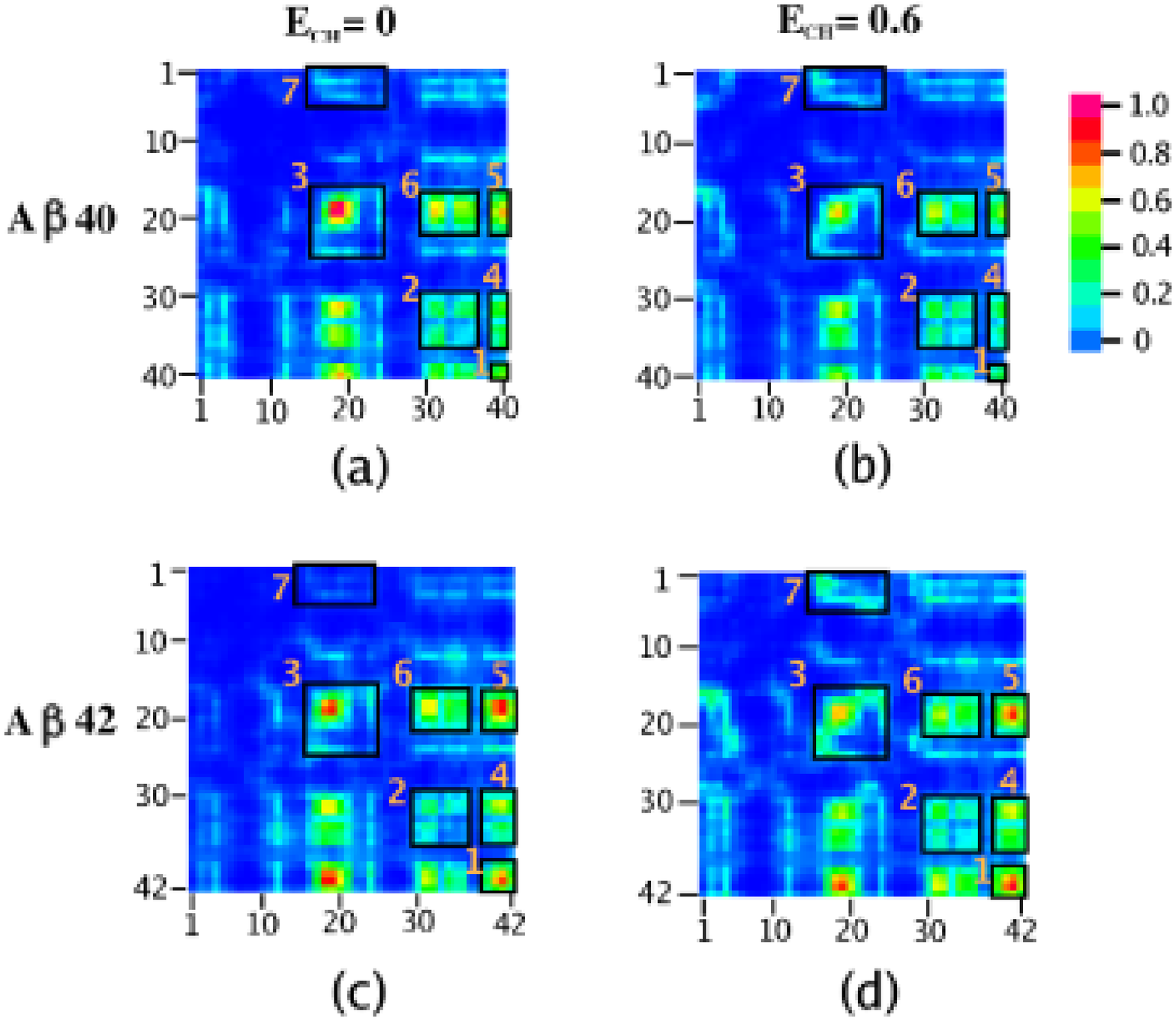}
\caption{}
\label{cm_pentamer_larger_inter}
 
\end{figure}

\newpage
\clearpage
\begin{figure}[ht]
\centering
\includegraphics*[width=14cm]{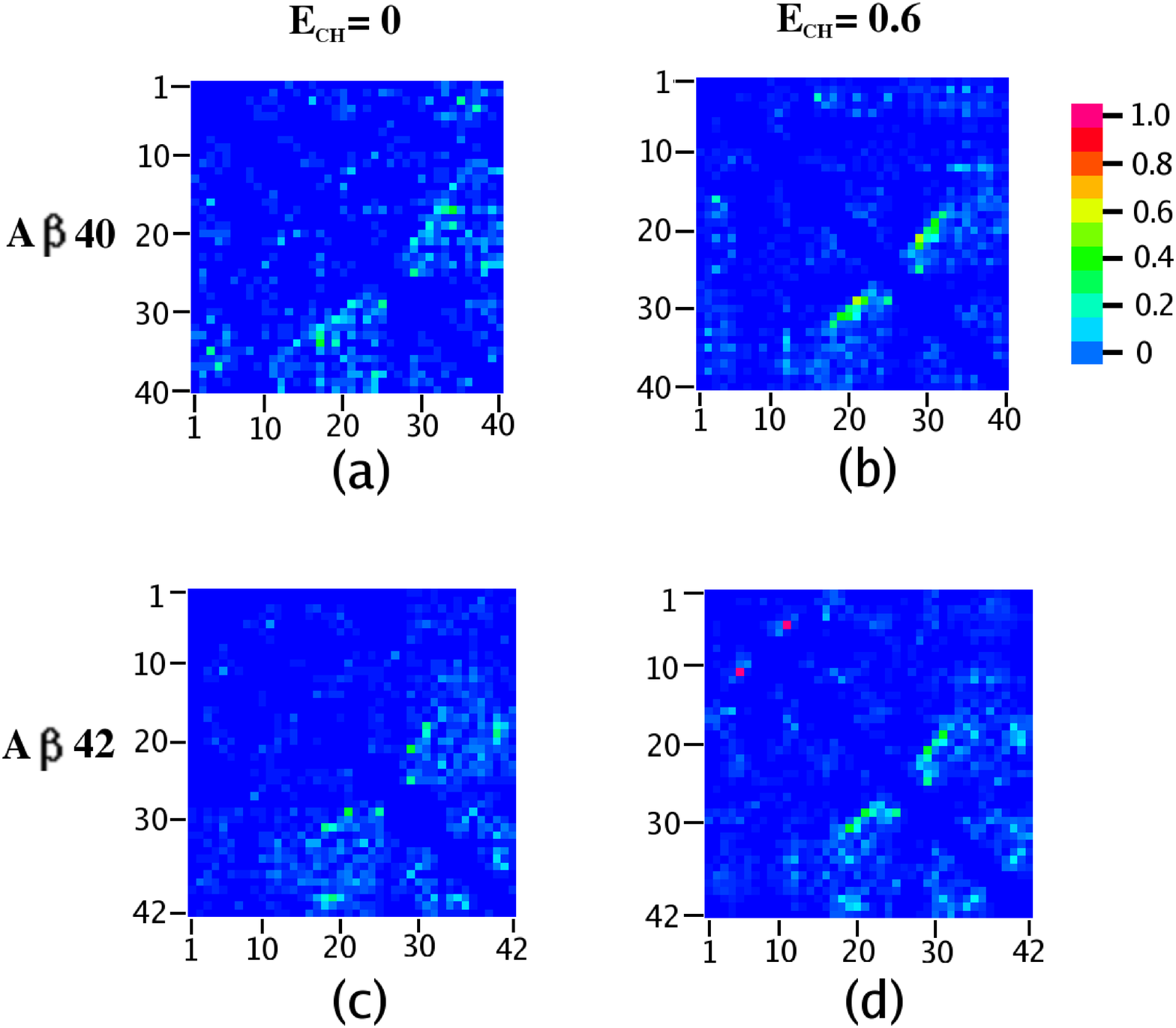}
\caption{}
\label{cm_hb_intra}
 
\end{figure}

\newpage
\clearpage
\begin{figure}[ht]
\centering
\includegraphics*[width=14cm]{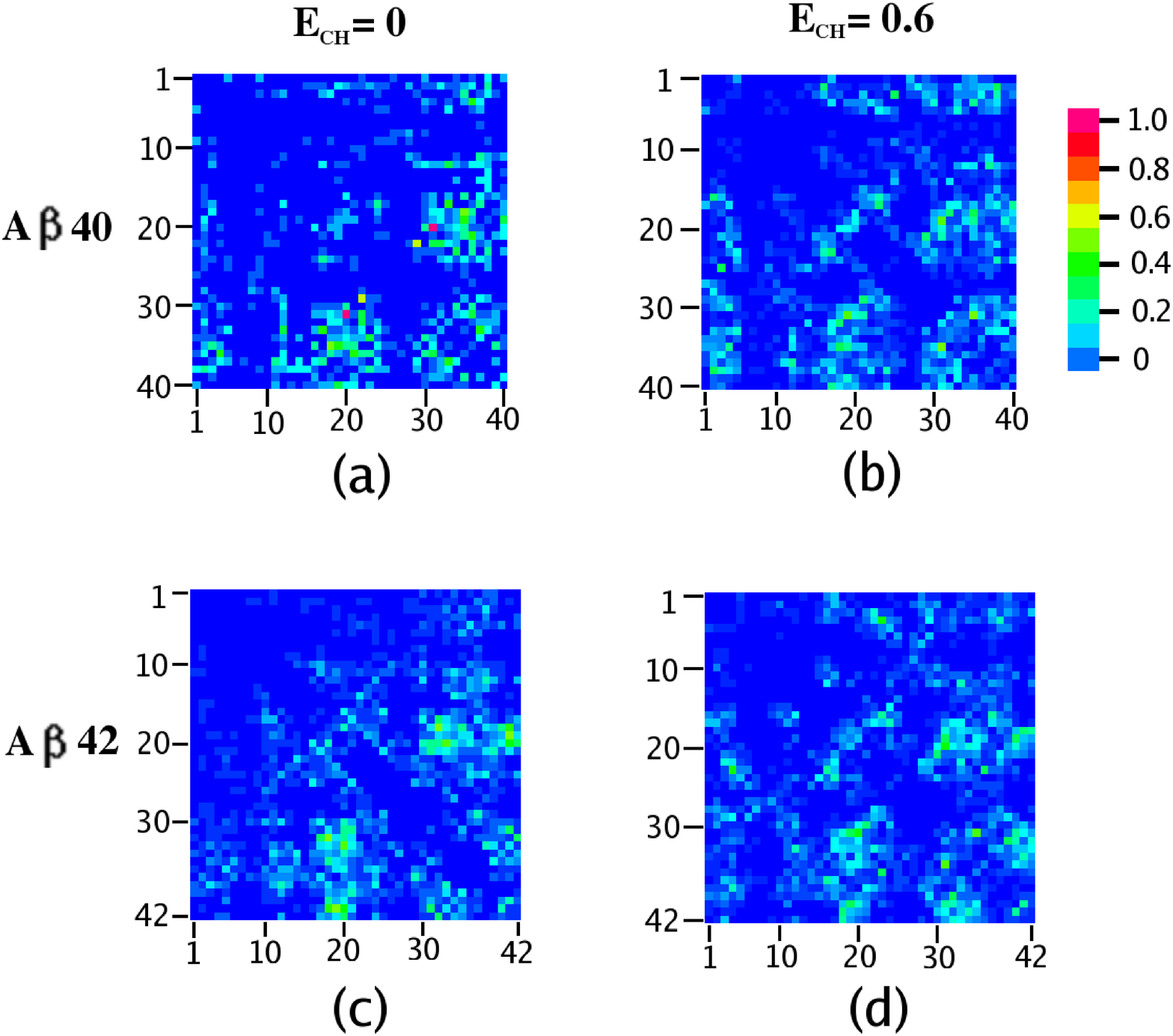}
\caption{}
\label{cm_hb_inter}
 
\end{figure}

\newpage
\clearpage
\begin{table}[ht]
\centering
\begin{tabular}{||c||c|c||c|c||}  \hline
                  &  \multicolumn{2}{c||}{A$\beta$40}      & \multicolumn{2}{c||}{A$\beta$42}       \\ \hline
                  &  $E_{CH}=0$        & $E_{CH}=0.6$       & $E_{CH}=0$          & $E_{CH}=0.6$     \\ \hline 
  \ Turn \           &\ 0.44 $\pm$ 0.04 \ &\ 0.43 $\pm$ 0.04 \  &\ 0.48 $\pm$ 0.04 \  &\ 0.50 $\pm$ 0.05 \ \\ \hline
  \ $\beta$-strand \ &\ 0.11 $\pm$ 0.02 \ &\ 0.12 $\pm$ 0.03 \  &\ 0.11 $\pm$ 0.03 \  &\ 0.10 $\pm$ 0.03 \ \\ \hline

\end{tabular}
\caption{}
\label{table_monomer_ss_average}
\end{table}

\newpage
\clearpage
\begin{table}[ht]
\centering
\begin{tabular}{||c||c|c||c|c||}  \hline
                  &  \multicolumn{2}{c||}{A$\beta$40}      & \multicolumn{2}{c||}{A$\beta$42}       \\ \hline
                  &  $E_{CH}=0$        & $E_{CH}=0.6$       & $E_{CH}=0$          & $E_{CH}=0.6$     \\ \hline 
  \ Turn \            &  \ 0.44 $\pm$ 0.02 \  & \ 0.45 $\pm$ 0.02 \  & \ 0.42 $\pm$ 0.02 \ & \ 0.45 $\pm$ 0.02 \ \\ \hline
  \ $\beta$-strand \  &  \ 0.13 $\pm$ 0.02 \  & \ 0.11 $\pm$ 0.01 \  & \ 0.13 $\pm$ 0.01 \ & \ 0.11 $\pm$ 0.01 \ \\ \hline

\end{tabular}
\caption{}
\label{table_pentamer_larger_ss_average}
\end{table}

\newpage
\clearpage
\begin{table}[ht]
\begin{tabular}{||c|c|c|c||c|c|c|c||} \hline \hline
  \multicolumn{4}{||c||}{A$\beta$40} &  \multicolumn{4}{c||}{A$\beta$42}  \\ \hline
  \multicolumn{2}{||c|}{$E_{CH}=0$}  & \multicolumn{2}{c||}{$E_{CH}=0.6$} 
    & \multicolumn{2}{c|}{$E_{CH}=0$} & \multicolumn{2}{c||}{$E_{CH}=0.6$}   \\ \hline

  L17 & 0.17   & M35 & 0.14   & V40 & 0.17   & A30 & 0.17  \\ \hline 
  A21 & 0.15   & G38 & 0.14   & I31 & 0.16   & G29 & 0.15  \\ \hline 
  G33 & 0.15   & I31 & 0.13   & G38 & 0.15   & E11 & 0.15  \\ \hline
  V24 & 0.14   & G33 & 0.12   & A21 & 0.13   & R5  & 0.14  \\ \hline 
  \ G38 \  & \ 0.14 \  & \ G37 \ & \ 0.12 \  & \ G29 \ & \ 0.13 \   & \ I31 \  & \ 0.14 \    \\ \hline \hline 
\end{tabular}
\caption{}
 \label{table_HB_intra_cumulative_pentamer_larger}
 
\end{table}

\newpage
\clearpage
\begin{table}[ht]
\begin{tabular}{||c|c|c|c||c|c|c|c||} \hline \hline
  \multicolumn{4}{||c||}{A$\beta$40} &  \multicolumn{4}{c||}{A$\beta$42}  \\ \hline
  \multicolumn{2}{||c|}{$E_{CH}=0$}  & \multicolumn{2}{c||}{$E_{CH}=0.6$} 
    & \multicolumn{2}{c|}{$E_{CH}=0$} & \multicolumn{2}{c||}{$E_{CH}=0.6$}   \\ \hline
  
  M35 & 0.15   & M35 & 0.14   & F20 & 0.18   & I31 & 0.14   \\ \hline
  I31 & 0.14   & G38 & 0.14   & V18 & 0.15   & G33 & 0.13   \\ \hline
  V36 & 0.14   & I31 & 0.13   & G33 & 0.13   & V40 & 0.13   \\ \hline
  G37 & 0.14   & G33 & 0.122   & I41 & 0.13   & L17 & 0.12   \\ \hline
 \  G38 \ & \ 0.13 \   & \ G37 \ & \ 0.12 \   & \ A21 \ & \ 0.12 \   & \ F20 \ & \ 0.12 \   \\ \hline \hline
\end{tabular}
\caption{}
 \label{table_HB_inter_cumulative_pentamer_larger}
 
\end{table}

\end{document}